\documentclass[aps,prb,10pt,twocolumn,superscriptaddress,floatfix,letter,longbibliography]{revtex4-2}
\usepackage{amsmath}
\usepackage{mathtools}
\usepackage{bbold}
\usepackage{mathrsfs}

\usepackage{graphicx}
\usepackage[caption=false]{subfig}
\usepackage{epstopdf} 
\usepackage{wrapfig}
\usepackage{braket}
\usepackage{xcolor}
\usepackage{comment}
\usepackage{graphicx}
\usepackage{lipsum}
\usepackage{cancel}

\usepackage{tikz}
\usepackage{circuitikz}
\usetikzlibrary{chains,fit,shapes}
\usetikzlibrary{cd} 
\usetikzlibrary{shapes.geometric}
\usetikzlibrary{decorations.markings}
\usetikzlibrary{decorations.pathmorphing,patterns}
\usetikzlibrary{decorations}
\usetikzlibrary{calc}
\usetikzlibrary{intersections}
\usetikzlibrary{arrows}
\usetikzlibrary{matrix}
\usetikzlibrary{positioning}
\usetikzlibrary{automata}
\newcommand{\RNum}[1]{\uppercase\expandafter{\romannumeral #1\relax}}

\newcommand{\btheta}{\boldsymbol{\theta}}

\newcommand{\Trace}{{\mathrm{Tr}}}

\newcommand{\diag}{{\mathrm{diag}}}

\def\ket#1{| #1\rangle}
\newcommand{\nep}{\mathrm{e}}

\newcommand{\QAOA}{\mathrm{\scriptscriptstyle QAOA}}

\newcommand{\x}{{\mathbf x}}

\newcommand{\A}{{\mathbf A}}
\newcommand{\B}{{\mathbf B}}

\newcommand{\rmC}{{\mathrm C}}

\newcommand{\target}{\mathrm{\scriptscriptstyle T}}
\newcommand{\starget}{s_{\target}}

\newcommand{\mix}{\mathrm{\scriptscriptstyle M}}

\newcommand{\Heis}{\mathrm{\scriptscriptstyle H}}

\newcommand{\transpose}{\mathsf{T}}

\newcommand{\Nbasis}{\mathrm{N_c}}

\newcommand{\bfP}{{\mathbf P}}
\newcommand{\W}{{\mathbf W}}
\newcommand{\bfK}{{\mathbf K}}
\newcommand{\Q}{{\mathbf Q}}
\newcommand{\U}{{\mathbf U}}
\newcommand{\V}{{\mathbf V}}

\newcommand{\Tr}{{\rm Tr}}

\newcommand{\prm}{\mathrm{p}}

\newcommand{\opc}[1]{{\hat{c}^{\phantom \dagger}}_{#1}}
\newcommand{\opcH}[1]{{\hat{c}^{\phantom \dagger\text{H}}}_{#1}}
\newcommand{\opcdag}[1]{{\hat{c}^{\dagger}}_{#1}}
\newcommand{\opcdagH}[1]{{\hat{c}^{\dagger\text{H}}}_{#1}}

\newcommand{\PauliSigma}{\hat{\sigma}}

\newcommand{\Ho}{\hat{H}}
\newcommand{\Uo}{\hat{U}}

\newcommand{\Nsites}{\mathrm{N}}
\newcommand{\Nfermions}{\mathrm{\hat{N}_F}}
\newcommand{\res}{\mathrm{res}}

\newcommand{\Ptrot}{\mathrm{P}}

\newcommand{\commentra}[1]
{{\textcolor{orange}{#1}}}

\newcommand{\mathH}{\mathbb{H}}
\newcommand{\mathGamma}{\mathbb{\Gamma}}
\newcommand{\mathW}{\mathbb{W}}
\newcommand{\mathP}{\mathbb{P}}
\newcommand{\opgamma}[1]{{\hat{\gamma}^{\phantom \dagger}}_{#1}}
\newcommand{\opgammadag}[1]{{\hat{\gamma}^{\dagger}}_{#1}}

\newcommand{\opbfPsi}[1]{{\widehat{\mathbf{\Psi}}^{\phantom \dagger}}_{#1}}
\newcommand{\opbfPsiH}[1]{{\widehat{\mathbf{\Psi}}^{\text{H}}}_{#1}}

\newcommand{\opbfPsik}[1]{{\widehat{\mathbf{\Psi}}^{\phantom \dagger }}_{k#1}}
\newcommand{\opbfPsiHk}[1]{{\widehat{\mathbf{\Psi}}^{\phantom \dagger \text{H}}}_{k#1}}
\newcommand{\opbfPsidag}[1]{{\widehat{\mathbf{\Psi}}^{\dagger}}_{#1}}
\newcommand{\opbfPsidagj}[1]{{\widehat{\mathbf{\Psi}}^{\dagger}}_{j#1}}
\newcommand{\opbfPsidagHj}[1]{{\widehat{\mathbf{\Psi}}^{\dagger \text{H}}}_{j#1}}

\newcommand{\opbfPhi}[1]{{\widehat{\mathbf{\Phi}}^{\phantom \dagger}}_{#1}}
\newcommand{\opbfPhik}[1]{{\widehat{\mathbf{\Phi}}^{\phantom \dagger}}_{k#1}}
\newcommand{\opbfPhidag}[1]{{\widehat{\mathbf{\Phi}}^{\dagger}}_{#1}}
\newcommand{\opbfPhidagj}[1]{{\widehat{\mathbf{\Phi}}^{\dagger}}_{j#1}}

\newcommand{\mathU}{\mathbb{U}}
\newcommand{\gs}{\mathrm{\scriptstyle gs}}

\graphicspath{{./}{./Fig/}}

\begin{document}
\author{Vincenzo Roberto Arezzo}
\affiliation{SISSA, Via Bonomea 265, I-34136 Trieste, Italy}

\author{Ruiyi Wang}
\affiliation{SISSA, Via Bonomea 265, I-34136 Trieste, Italy}

\author{Kiran Thengil}
\affiliation{SISSA, Via Bonomea 265, I-34136 Trieste, Italy}

\author{Giovanni Pecci}
\affiliation{SISSA, Via Bonomea 265, I-34136 Trieste, Italy}

\author{Giuseppe E. Santoro}
\affiliation{SISSA, Via Bonomea 265, I-34136 Trieste, Italy}
\affiliation{CNR-IOM - Istituto Officina dei Materiali, Consiglio Nazionale delle Ricerche, c/o SISSA Via Bonomea 265, 34136 Trieste, Italy}
\affiliation{International Centre for Theoretical Physics (ICTP), P.O.Box 586, I-34014 Trieste, Italy}

\title{Digital controllability of transverse field Ising chains}

\begin{abstract}
Quantum Annealing (QA) encounters limitations when the energy gap along the annealing path becomes exponentially small, leading to impractically long runtimes. 
In contrast, the success of hybrid digital methods like the Quantum Approximate Optimization Algorithm (QAOA), which operate via discrete unitary operations, relies on the optimization of the variational parameters appearing in the state. 
We analyze a class of transverse-field Ising models which includes problems with exponentially small spectral gaps, but whose dynamics is described in terms of fermionic Gaussian states after Jordan-Wigner mapping. 
We show that, for digital alternating QAOA-like states, the number of unitaries required to reach the exact ground state scales quadratically with system size and is independent of the annealing gap. This number can be exactly computed from the algebraic properties of the {\em Ansatz}, revealing a fundamental distinction between digital methods and their analog counterpart.
\end{abstract}

\maketitle

\section{Introduction}
Quantum annealing (QA) is one of the prominent algorithms for quantum optimization and ground state preparation \cite{finnila_quantum_1994,kadowaki1998quantum,Santoro_SCI02,Santoro_2006,Albash_RMP18}. It relies on the quantum evolution of the ground state of a simple Hamiltonian --- referred to as mixer or driving Hamiltonian --- towards the ground state of a target Hamiltonian, encoding the solution of a given optimization problem. If the evolution is slow enough, the adiabatic theorem ensures that the system remains in the instantaneous ground state throughout the whole dynamics. However, the performance of QA is severely influenced by the instantaneous spectral gap of the Hamiltonian driving the dynamics: when the lowest spectral gap narrows with increasing system size, the annealing time required to successfully maintain adiabaticity, reaching the target state, grows correspondingly. 
Among the techniques to avoid this bottleneck, the design of optimal annealing schedules interpolating between the mixer and the target Hamiltonians has been widely pursued~\cite{Matsuura_PRA2021,Cote_2023,Quiroz_PRA2019,Lucignano_PRA2022,grabarits2025nonadiabatic,Passarelli_PRB2019,Caneva_PRA2011,Montangero_PRA2011,Montangero_dCRAB_PRA2015}; alternatively, the addition of counter-diabatic terms to suppress the transitions caused by the closing gap has also been investigated~\cite{Berry_2009, KOLODRUBETZ20171, Wurtz2022counterdiabaticity, balducci2024fighting}.

As an alternative to QA, a class of Variational Quantum Algorithms (VQAs), based on parametric quantum circuits, has been developed for quantum optimization (see Ref.~\cite{cerezo_variational_2021} and references therein). These VQA algorithms share conceptual similarities with QA: starting from an easy-to-prepare initial state, the system is evolved through the application of parametrized quantum gates. The gate parameters are then optimized to minimize the expectation value of the target Hamiltonian on the final state. 
Upon convergence, the final state approximates the ground state of the target Hamiltonian. One of the most studied VQA is the Quantum Approximate Optimization Algorithm (QAOA) \cite{farhi_quantum_2014}. 
It consists in alternating unitary transformations generated by the target, $\Ho_{\target}$, and mixer, $\Ho_{\mix}$, Hamiltonians
\begin{equation} \label{eqn:QAOA_state_intro}
|\psi(\btheta)\rangle =
\nep^{-i\theta^{\mix}_{\Ptrot} \Ho_{\mix}} 
\nep^{-i\theta^{\target}_{\Ptrot} \Ho_{\target}} \cdots
\nep^{-i\theta^{\mix}_{1} \Ho_{\mix}} 
\nep^{-i\theta^{\target}_{1} \Ho_{\target}} 
 |\psi_0\rangle \;,
 \nonumber
\end{equation}
starting from the ground state $|\psi_0\rangle$ of the mixer $\Ho_{\mix}$.
Each of the $\Ptrot$ layers of the circuit is parametrized by $2$ parameters, for a total of $2\Ptrot$ variational parameters to optimize at each run. 
%
%
%
In the adiabatic regime, i.e., for very small parameters, QAOA closely approximates QA and succeeds when the circuit depth is comparable to the annealing time of the corresponding analog problem \cite{farhi_quantum_2014, zhou_quantum_2020}. 
In this limit, it inherits the same limitations as QA, in particular the poor performance associated to exponentially closing spectral gap as the system size increases. 
However, unlike traditional QA (or Adiabatic Quantum Computation~\cite{Albash_RMP18}), QAOA does not necessarily rely on adiabatic evolution.
The optimized digital dynamics potentially exploits diabatic processes to better approximate the target state, circumventing the small gap bottleneck. 
Consequently, for general non-adiabatic digital protocols, how the QAOA circuit depth must scale with system size to {\em exactly} reach the target state remains an open problem.

We focus particularly on frustrated transverse field Ising chains that exhibit exponentially small spectral gaps as the number of spins increases
~\cite{Knysh_PRA2020, Cote_2023, wang2025exponentialquadraticoptimalcontrol}.
The fact that smooth optimal QAOA schedules circumventing the small gap bottleneck problem can be constructed by adapting Chopped Random Basis (CRAB) ~\cite{Montangero_PRA2011,Montangero_dCRAB_PRA2015} techniques of optimal control to the digital case was already demonstrated in Ref.~\cite{wang2025exponentialquadraticoptimalcontrol}.
Here we 
show that, for all the models in this class, the critical QAOA depth $\Ptrot^{\mathrm{cr}}_{\Nsites}$ beyond which the {\em exact} ground state is reached scales {\em quadratically} with the system size $\Nsites$, regardless of the exponentially small spectral gap of the corresponding annealing Hamiltonian.
This $\Ptrot^{\mathrm{cr}}_{\Nsites}$ depth correspond exactly with the dimension of the manifold of Gaussian fermionic states.
This result marks a fundamental difference between digital quantum computation and adiabatic quantum computation in continuous time: while the latter is limited by the spectral gap, the success of the digital approach in the non-adiabatic regime is only restricted by purely algebraic constraints.

Our paper is organized as follows. 
Section \ref{sec:models} introduces the theoretical methods we use, in particular, Sec.~\ref{sec:BdG_ph}: the Bogoljubov-de Gennes theory for the digital case, Sec.~\ref{sec:frustrated}: the models considered, Sec.~\ref{sec:reflection_symmetry}: the role of spatial symmetry.
Section \ref{sec:numerical} presents the numerical results corroborating our theoretical results for $\Ptrot^{\mathrm{cr}}_{\Nsites}$.
Section \ref{sec:summary}, finally, summarizes our results and discusses possible perspectives of our work.
Technical aspects of our work are contained in the final Appendices. 

\section{Models and Methods} \label{sec:models}
Consider a quantum Ising chain Hamiltonian
\begin{equation}
\label{eqn:isingmodelham}
    \Ho(s) = s\Ho_z + (1-s) \Ho_x \;,
\end{equation}
where
\begin{equation}
\label{eqn:isingmodels}
    \Ho_{z} = -\sum_{j=1}^\Nsites J_j \PauliSigma^z_j\PauliSigma^z_{j+1} \;, \hspace{5mm} 
    \Ho_{x} = - h \sum_{j=1}^\Nsites\PauliSigma^x_j \;.
\end{equation}
As written, this is a convex combination of the classical interaction term $\Ho_z$, with couplings $J_j$, and transverse field term $\Ho_x$.
For $s=0$, the ground state of $\Ho(s)$ is that of $\Ho_x$, i.e., (for $h>0$) the product state $|\psi_0\rangle = \ket{+}^{\otimes \Nsites}$, where $\ket{+}=\frac{1}{\sqrt{2}}(\ket{\!\uparrow}+\ket{\!\downarrow})$ is the eigenstate of $\PauliSigma^x$ with eigenvalue $+1$.

Suppose that our goal is to reach the ground state of $\Ho(\starget)$ with a given target value $\starget\in (0,1]$. 
In a traditional QA approach, we would use a linear-schedule $s(t)=(t/\tau)\starget$, where $\tau$ is a large annealing time, and adiabatically evolve the system according to the Schr\"odinger dynamics, from time $t=0$ to $t=\tau$.
As is well known, the annealing time needed to reach the target state while remaining adiabatically close to the instantaneous ground state of $\Ho(s)$ is $\tau_{\text{ad}} \propto \Delta_{\min}^{-2}$, where $\Delta_{\min}$ is the minimum instantaneous spectral gap between the ground and the first excited state of $\Ho(s)$.

In QAOA~\cite{farhi_quantum_2014}, one relies on a quantum circuit of depth $\Ptrot$:
\begin{equation} \label{eqn:QAOA_state}
|\psi_{\Ptrot}^{\QAOA}(\btheta)\rangle =
\Uo_{\Ptrot}(\theta^x_\Ptrot,\theta^z_\Ptrot) \, 
\cdots 
\Uo_1(\theta^x_1,\theta^z_1) \, |\psi_0\rangle \;,
\end{equation}
where
\begin{equation} \label{eqn:1st_Trotter}
\Uo_p(\theta^x_p,\theta^z_p)=\nep^{-i\theta^x_p \Ho_x} 
\nep^{-i\theta^z_p \Ho_z}  \;,
\end{equation}
with $p=1,...,\Ptrot$. 
To ensure the convergence to the ground state of $\Ho(\starget)$, we optimize the $2\Ptrot$ parameters $\btheta = (\theta^x_1 \dots \theta^x_\Ptrot; \theta^z_1 \dots \theta^z_\Ptrot)$ in order to minimize the mean final energy:
\begin{equation}\label{eqn:qaoa_energy}
   E_{\Ptrot}(\starget) \equiv \braket{\psi_{\Ptrot}^{\QAOA}(\btheta) | \Ho(\starget) | \psi_{\Ptrot}^{\QAOA}(\btheta)} \;.
\end{equation}

The case of a transverse field quantum Ising chain is very special, as we can apply a Jordan-Wigner transformation to map the problem to a quadratic fermionic model~\cite{mbeng2024quantum}. 
This transformation projects the evolution on the manifold of fermionic Gaussian states, and allows us to describe the digital quantum dynamics using Bogoljubov-de Gennes equations.

In particular, we implement the Nambu formalism \cite{mbeng2024quantum} and cast the fermionic Hamiltonians $\Ho_z$ and $\Ho_x$ in matrix form:
\begin{eqnarray} \label{quadratic-H:eqn}
\Ho_{x/z} = \opbfPsidag{} \, \mathH_{x/z} \, \opbfPsi{} \;,
\end{eqnarray}
where $\opbfPsi{} = (\opc{1}, \dots, \opc{\Nsites}, \opcdag{1}, \dots, \opcdag{\Nsites})^\transpose$ is a $2\Nsites$-dimensional Nambu vector combining fermionic annihilation ($\opc{j}$) and creation ($\opcdag{j}$) operators, and $\mathH_{x/z}$ are $2\Nsites\times 2\Nsites$ Hermitian matrices.
For the transverse field part we obtain
\begin{equation}
    \mathH_x = 
    \left( \begin{array}{cc} \A_x & \mathbf{0} \\
                           \mathbf{0} & -\A_x \end{array} \right) \;,
\end{equation}
where $\A_x$ is diagonal and proportional to the identity $\A_x = h \mathbf{1}$.
For the $\Ho_z$ coupling term we get
\begin{equation}
    \mathH_z = 
    \left( \begin{array}{cc} \A_z & \B_z \\
                           -\B_z & -\A_z \end{array} \right) \;,
\end{equation}
where $\A_z$ and $\B_z$ are $\Nsites\times \Nsites$ real  symmetric or anti-symmetric matrices, respectively, and their only non-zero elements are given by:
\begin{equation} \label{eqn:AB_z}
\left\{
\begin{array}{l}
(\A_z)_{j,j+1} = (\A_z)_{j+1,j} = -\displaystyle
J_j/2 \vspace{4mm} \\
(\B_z)_{j,j+1} = -(\B_z)_{j+1,j} = - \displaystyle
J_j/2
\end{array}
\right. \;,
\end{equation}
with additional matrix elements, dependent on the fermionic parity, given by:
\begin{equation}
\label{eqn:A_zborder}
(\A_z)_{\Nsites,1} = (\A_z)_{1,\Nsites} = (-1)^{\prm} J_{\Nsites}/2\;,  
\end{equation}
and
\begin{equation}
\label{eqn:Bzborder}
(\B_z)_{\Nsites,1} = -(\B_z)_{1,\Nsites} = (-1)^{\prm} J_{\Nsites}/2 
\;.
\end{equation}
Here $\prm=0$ or $1$ denotes even or odd fermion parity sectors of the Hilbert space: $(-1)^{\prm}=\nep^{i\pi \Nfermions}$, where $\Nfermions=\sum_j \opcdag{j}\opc{j}$ is the total number of fermions.
Finally, the target Hamiltonian is given by $\Ho(\starget) = \opbfPsidag{} \,\mathH(\starget) \opbfPsi{}$, where 
\begin{equation}
\mathH(\starget) = \starget\, \mathH_z + (1-\starget)\, \mathH_x \;.
\end{equation}

In order to proceed, it is convenient to express the QAOA dynamics in the Heisenberg picture.
The digitized dynamics implied by the state evolution in Eq.~\eqref{eqn:QAOA_state} is captured by the Bogoljubov-de Gennes equations for the evolution operator~\cite{mbeng2024quantum}, see Appendix \ref{app:Heisenberg_qaoa_dynamics} for details, represented by a $2\Nsites\times 2\Nsites$ unitary matrix
\begin{equation}
\label{eq:Uoperator}
    \mathU(\btheta) =
\mathU_\Ptrot(\theta^x_\Ptrot,\theta^z_\Ptrot) \, \cdots \,
    \mathU_1(\theta^x_1,\theta^z_1)\, \mathU_0 \;,
\end{equation}
with
\begin{equation} \label{eqn:U_nambu}
\mathU_p(\theta^x_p,\theta^z_p) =
\nep^{-2i \theta^x_p \mathH_x} 
\nep^{-2i \theta^z_p \mathH_z} \;.
\end{equation}
The unitary matrix $\mathU_0$ encodes the transformation from the original fermions $\opbfPsi{}\!\!$ to the Bogoljubov fermions for which the initial state $|\psi_0\rangle$ is a vacuum, see Appendix \ref{app:Heisenberg_qaoa_dynamics}. 
For $h>0$, we have $\mathU_0=\mathbb{1}$, 
while for $h<0$
\begin{equation}
\mathU_0 = 
\left( \begin{array}{cc} 
\mathbf{0}_\Nsites & \mathbf{1}_\Nsites \\
\mathbf{1}_\Nsites & \mathbf{0}_\Nsites \end{array}   \right) \;.
\end{equation}
The mean final energy, see Appendix \ref{app:Heisenberg_qaoa_dynamics} for a derivation, reads
\begin{equation} \label{eqn:nambu_qaoa_energy}
    E_{\Ptrot}(\starget) = \Trace\left(\mathU^{\dagger}(\btheta) \, \mathH(\starget) \, \mathU(\btheta) \, \mathGamma\right) \;,
\end{equation}
where $\mathGamma$ is a $2\Nsites\times 2\Nsites$ matrix given by
\begin{equation}
\mathGamma = \left( 
    \begin{array}{cc}  \mathbf{0}_\Nsites & \mathbf{0}_\Nsites \\
    \mathbf{0}_\Nsites & \mathbf{1}_\Nsites 
    \end{array}
            \right) \;.
\end{equation}

Consider now the unitary Bogoljubov transformation $\mathU_{\target}$ that diagonalizes $\mathH(\starget)$, 
\begin{equation}
    \mathU^{\dagger}_{\target} \, \mathH(\starget) \, \mathU_{\target} \equiv \mathH_{\text{d}}(\starget) =  \diag(\epsilon_1,\cdots,\epsilon_{\Nsites},-\epsilon_1,\cdots,-\epsilon_{\Nsites}) \nonumber
\end{equation}
with its particle-hole symmetric eigenvalues $\pm \epsilon_j$,
ordered in such a way that $E_{\target}=-\sum_j \epsilon_j$ is the ground state energy of the target Hamiltonian $\Ho_{\target}=\Ho(s_{\target})$. 
Then, if $\mathU(\btheta)$ coincides with $\mathU_{\target}$, it follows from Eq.~\eqref{eqn:nambu_qaoa_energy} that 
$E_{\Ptrot}(\starget)=E_{\target}=-\sum_j \epsilon_j$: the target state is obtained. 
(Usually, the occupied single-particle states have all negative energies, i.e., $\epsilon_j>0$, but this is not always the case, as explained in Ref.~\cite{wang2025exponentialquadraticoptimalcontrol}.)

It turns out that there is a ``gauge freedom'' related to mixing the $\Nsites$ Bogoliubov modes which are ``occupied'' in the ground state, leaving the final energy Eq.~\eqref{eqn:nambu_qaoa_energy} invariant. 
This mixing is realized by a unitary matrix of the form
\begin{equation} \label{eqn:W}
\mathW = \left( 
\begin{array}{cc}   
\mathbf{W} & \mathbf{0} \\
\mathbf{0} & \mathbf{W^*} 
\end{array} \right)    \;,
\end{equation}
for any $\mathbf{W}\in U(\Nsites)$, which commutes with $\mathGamma$ due to its block structure, 
\begin{equation} \label{eqn:gauge_commut}
    [\mathGamma,\mathW]=0 \;.
\end{equation} 
In essence, it is sufficient that $\mathU(\btheta)$ coincides with $\mathU_{\target}$ up to a mixing of occupied states performed by $\mathW$, i.e., $\mathU(\btheta)=\mathU_{\target} \mathW$.
This follows immediately from Eq.~\eqref{eqn:nambu_qaoa_energy}
%
%
by using Eq.~\eqref{eqn:gauge_commut} and the cyclic property of the trace:
\begin{align}
\label{eqn:unaltereddenergy}
E_{\Ptrot}(\starget) &= \Tr\left(\mathW^{\dagger} \mathU^{\dagger}_{\target} \,\mathH(\starget) \, \mathU_{\target} \mathW \, \mathGamma \right) \nonumber \\ 
&= \Tr\left( \mathU^{\dagger}_{\target} \, \mathH(\starget) \, \mathU_{\target} \, \mathGamma \right)  
= E_{\target} \;.
\end{align}

Summarizing, to \emph{exactly} solve the optimization problem, we should find a parameter vector $\btheta$ such that $\mathU(\btheta) \equiv \mathU_{\target} \mathW$.
Let us denote by $\dim_U$ the (real) dimensionality of the group to which $\mathU_{\target}$ belongs and by 
$\dim_W$ the dimensionality of the ``gauge freedom'' block matrix $\mathW$, which in absence of any additional symmetry is given by $\dim_W=\Nsites^2$. 
The dimensionality of the QAOA unitary $\mathU(\btheta)$ is given by $2\Ptrot$, the total number of variational parameters. To solve the optimization problem, the matching of dimensionalities requires that:
\begin{equation}
2 \Ptrot \ge \dim_U - \dim_W .
\end{equation}


In the next section, we will calculate the dimension $\dim_U$ of $\mathU_{\target}$ by enforcing the particle-hole symmetry intrinsic in the Bogoljubov-de Gennes theory. Later on, we will discuss the modifications induced by the possible presence or absence of spatial symmetries.  

\subsection{Particle-Hole symmetry in the Bogoljubov-de Gennes dynamics} \label{sec:BdG_ph}
The QAOA evolution preserves all the symmetries of the Nambu Hamiltonians $\mathH_x$ and $\mathH_z$. 
%
In particular, the Nambu formalism inherently implements particle-hole symmetry, which implies that the evolution operator we aim at building will have the special form~\cite{mbeng2024quantum}:
\begin{equation} \label{eqn:U_nambu_UV}
\mathU = \left( 
\begin{array}{cc}  
\mathbf{U} & \mathbf{V^*} \\
\mathbf{V} & \mathbf{U^*}
\end{array}
\right) \;.
\end{equation}

Let us start by calculating the dimensionality of the group of such particle-hole symmetric unitary matrices. The fact that $\mathU^{\dagger}\mathU=\mathbb{1}$ imposes
the following constraints on $\mathbf{U}$ and $\mathbf{V}$:
\begin{equation} \label{eqn:ph-symmetry-constraint}
\begin{aligned}
\U^{\dagger}\U + \V^{\dagger}\V &= \mathbf{1}_\Nsites  \\
\U^{\dagger}\V^* + \V^\dagger\U^* &= \mathbf{0}_\Nsites
\end{aligned}
\;.
\end{equation}
The first requirement fixes a total of 
$\Nsites + 2\frac{\Nsites(\Nsites-1)}{2} = \Nsites^2$ real constraints, because $\U^{\dagger}\U + \V^{\dagger}\V$ is an $\Nsites\times \Nsites$ Hermitian. 
The second condition, instead, contributes $2(\Nsites + \frac{\Nsites(\Nsites-1)}{2}) = \Nsites^2 + \Nsites$ real constraints, because $\U^{\dagger}\V^* + \V^\dagger\U^*$ is a complex symmetric $\Nsites\times \Nsites$ matrix. 
Taken together, the particle-hole conditions implied by the special form of Eq.~\eqref{eqn:U_nambu_UV} give a total
of
\[ 
\textrm{N}_{\textrm{p-h constraints}} = 2\Nsites^2 + \Nsites \;,
\]
real constraints. 
%
To specify $\mathbf{U}$ and $\mathbf{V}$ we need a total of $4\Nsites^2$ real parameters. Therefore,
the subgroup of $U(2\Nsites)$ satisfying the p-h conditions in Eq.~\eqref{eqn:U_nambu_UV} has a dimensionality
\begin{align}
\dim_U &= 4\Nsites^2 - \textrm{N}_{\textrm{p-h constraints}} 
\nonumber \\
&= 4\Nsites^2 - (2\Nsites^2 + \Nsites)  = 2\Nsites^2 - \Nsites \;. 
\end{align}
Since $\W \in U(\Nsites)$, the dimension of the group generated by $\mathW$ is 
\[
\dim_W = \Nsites^2 \;,
\]
therefore knowing that the number of real parameters in the Bogoljubov-de Gennes QAOA form of $\mathU(\btheta)$ is $2\Ptrot$ we expect that exact $\mathU_{\target}$ can be in principle attained when: 
\begin{equation}
\label{eqn:inequalitydimension}
2\Ptrot \geq  \dim_U-\dim_W  
= \Nsites(\Nsites-1) \;.
\end{equation}
%

Interestingly, we can show directly that  $\dim_U-\dim_W =\Nsites(\Nsites-1)$ coincides with the dimension of the Gaussian fermionic state space $\dim_F$.
Indeed, if we consider a Gaussian state $\ket{\psi}$, we can always express it using \emph{Thouless formula} \cite{Thouless_NP1960}:
\begin{equation}
\label{eqn:gaussianstate}
    \ket{\psi_{\mathrm{Gaussian}}} = \mathcal{N}\exp\left(\frac{1}{2}\sum_{jj'}^{\Nsites}\mathbf{Z}_{jj'}\opcdag{j}\opcdag{j'}\right) \ket{0},
\end{equation}
where $\mathcal{N}$ is a normalization factor, $\mathbf{Z}_{jj'}$ is a $\Nsites \times \Nsites$ complex antisymmetric matrix \cite{Thouless_NP1960,mbeng2024quantum} that characterizes the state, and $\ket{0}$ is the vacuum of the $\opcdag{j}$ operators.
Since $\mathbf{Z}_{jj'}$ is complex antisymmetric, to characterize a generic Gaussian state $\ket{\psi_{\mathrm{Gaussian}}}$ we need $\Nsites(\Nsites-1)/2$ complex parameters, therefore
\begin{equation}
\label{eqn:dimensionequality}
\dim_{F} = \Nsites(\Nsites-1) \equiv \dim_{U} - \dim_{W} \;,
\end{equation}
real parameters.
If we take the QAOA ansatz state $\ket{\psi(\btheta)}$ in Eqs.~\eqref{eqn:QAOA_state},~\eqref{eqn:1st_Trotter}, which we know is a Gaussian state, having $2 \Ptrot > \dim_{F}$ parameters guarantees that the dimension of the QAOA {\em Ansatz} state is larger than the dimension of fermionic Gaussian states in which our target lives. 
%

We conjecture that
the critical depth of a QAOA circuit beyond which \textbf{controllability} is guaranteed, i.e., the target state can be reached, is the minimum bound of Eq.~\eqref{eqn:inequalitydimension}:
\begin{equation} \label{eqn:Pcr}
\Ptrot^{\mathrm{cr}}_{\Nsites} = \frac{\dim_{U} -\dim_{W}}{2} = \frac{\Nsites(\Nsites-1)}{2}  \;.
\end{equation}
This conjecture will be ``proved'' numerically in Sec. \ref{sec:numerical}.

\subsection{Frustrated Ising chain models}
\label{sec:frustrated}
So far, our arguments were general and apply to generic Ising model couplings $J_j$. 
We now further discuss the possible role that spatial symmetries play. But first, we specialize our discussion to a series of interesting special coupling cases, with and without disorder. 

We start considering the non-random case, in which there is no disorder in the couplings.

\subsubsection{Non-random case}
\begin{figure}[!htp]
\centering
\begin{tikzpicture}[scale=1.7,line cap=round,line width=2pt]
\filldraw [fill=black!0!white] (0,0) circle (2cm);
\draw [line width = 0.3mm, draw=blue, dashed] (-2.2,0) -- (2.0,0) node[right, black] {};
\foreach \x [evaluate=\x as \angle using (\x-0.5)*360/13] in {1,...,13}
{
\draw[line width=1pt,fill=white] ({2*cos(\angle)},{2*sin(\angle)}) circle (3mm);
}
\node[draw=none,font=\tiny,text=black,scale=1.5] at ({2*cos(0.5*360/13)},{2*sin(0.5*360/13)}) {$1$};
\node[draw=none,font=\tiny,text=black,scale=1.5] at ({2*cos(1.5*360/13)},{2*sin(1.5*360/13)}) {$2$};
\node[draw=none,font=\tiny,text=black,scale=1.5] at ({2*cos(2.5*360/13)},{2*sin(2.5*360/13)}) {$\cdots$};
\node[draw=none,font=\tiny,text=black,scale=1.5] at ({2*cos(4.5*360/13)},{2*sin(4.5*360/13)}) {$\cdots$};
\node[draw=none,font=\tiny,text=black,scale=1.5] at ({2*cos(5.5*360/13)},{2*sin(5.5*360/13)}) {$\scriptstyle{\frac{\Nsites-1}{2}}$};
\node[draw=none,font=\tiny,text=black,scale=1.5] at ({2*cos(6.5*360/13)},{2*sin(6.5*360/13)}) {$\scriptstyle{\frac{\Nsites+1}{2}}$};
\node[draw=none,font=\tiny,text=black,scale=1.5] at ({2*cos(7.5*360/13)},{2*sin(7.5*360/13)}) {$\scriptstyle{\frac{\Nsites+3}{2}}$};
\node[draw=none,font=\tiny,text=black,scale=1.5] at ({2*cos(8.5*360/13)},{2*sin(8.5*360/13)}) {$\cdots$};
\node[draw=none,font=\tiny,text=black,scale=1.5] at ({2*cos(-0.5*360/13)},{2*sin(-0.5*360/13)}) {$\Nsites$};
\node[draw=none,font=\tiny,text=black,scale=1.5] at ({2*cos(-1.5*360/13)},{2*sin(-1.5*360/13)}) {$\scriptstyle{\Nsites-1}$};
\node[draw=none,font=\tiny,text=black,scale=1.5] at ({2*cos(-2.5*360/13)},{2*sin(-2.5*360/13)}) {$\cdots$};
\node[draw=none,font=\tiny,text=black,scale=1.5] at ({2.25*cos(5*360/13)},{2.25*sin(5*360/13)}) {$J$};
\node[draw=none,font=\tiny,text=black,scale=1.5] at ({2.25*cos(6*360/13)},{2.25*sin(6*360/13)}) {$J_w'$};
\node[draw=none,font=\tiny,text=black,scale=1.5] at ({2.25*cos(7*360/13)},{2.25*sin(7*360/13)}) {$J_w$};
\node[draw=none,font=\tiny,text=black,scale=1.5] at ({2.25*cos(8*360/13)},{2.25*sin(8*360/13)}) {$J$};
\node[draw=none,font=\tiny,text=black,scale=1.5] at ({2.25*cos(2*360/13)},{2.25*sin(2*360/13)}) {$J$};
\node[draw=none,font=\tiny,text=black,scale=1.5] at ({2.25*cos(1*360/13)},{2.25*sin(1*360/13)}) {$J$};
\node[draw=none,font=\tiny,text=black,scale=1.5] at ({2.25*cos(0)},{2.25*sin(0)}) {$\scriptstyle{-J_f}$};
\node[draw=none,font=\tiny,text=black,scale=1.5] at ({2.25*cos(-1*360/13)},{2.25*sin(-1*360/13)}) {$J$};
\node[draw=none,font=\tiny,text=black,scale=1.5] at ({2.25*cos(-2*360/13)},{2.25*sin(-2*360/13)}) {$J$};
\end{tikzpicture}
\caption{The Ising frustrated ring model with an odd number of sites $\Nsites$. For $J_w'=J_w$ the model has a reflection symmetry around the axis indicated by the dashed blue line.}
\label{fig:ring}
\end{figure}
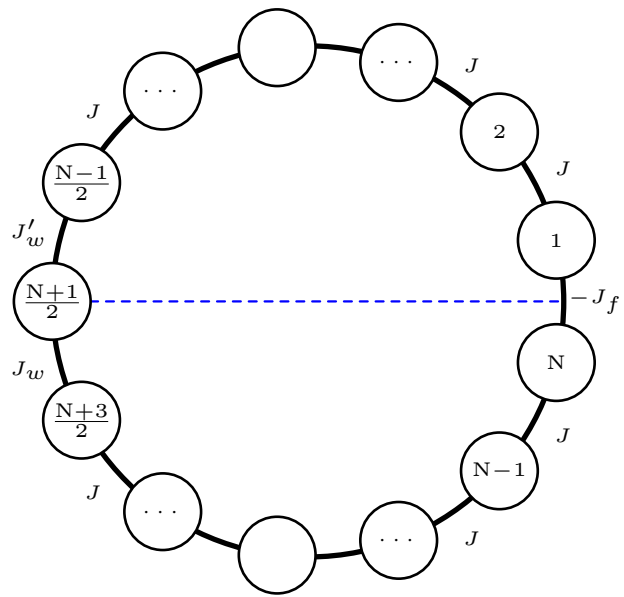

We take the model in Eqs.~\eqref{eqn:isingmodelham}-\eqref{eqn:isingmodels}, with the following couplings $J_j$:
\begin{equation}
  J_{j} =
    \begin{cases}
      J_w & \text{if $j=(\Nsites+ 1)/2$}\\
      J'_w & \text{if $j=(\Nsites- 1)/2$}\\
      -J_f & \text{if $j=\Nsites$}\\
      J & \text{otherwise}
    \end{cases}   \;,
\label{eq:couplings}
\end{equation}
sketched in Fig.~\ref{fig:ring}.

\begin{figure}[htp]
\includegraphics[width=\columnwidth]{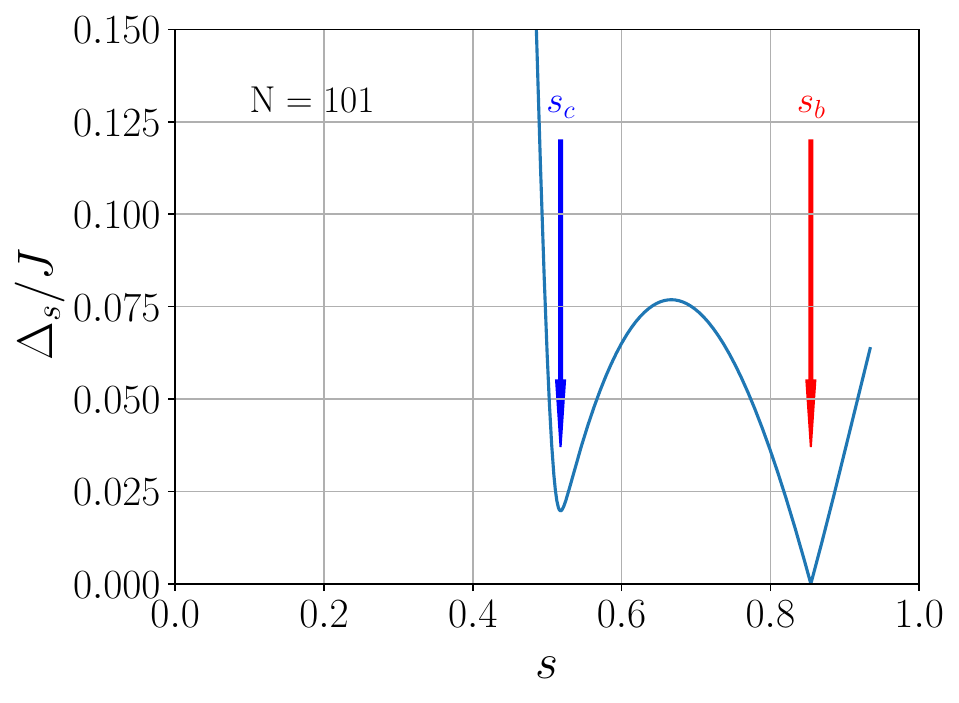}
\vspace{-5mm}
\caption{The spectral gap $\Delta_s$ between ground state and first excited state of a Ising frustrated ring model of size $\Nsites=101$ as a function of the interpolation parameter $s$, see Eq.~\eqref{eqn:isingmodelham}.
Here $J_w/J=0.5$, $J_w'/J=0.55$, $J_f/J=0.45$ and the spectral bottleneck occurs at $s_b\approx 0.8544$.
The corresponding result for $J_w'=J_w$ is given in Fig.~2 of Ref.~\cite{wang2025exponentialquadraticoptimalcontrol}. 
}
\label{fig:gap_N101_unsymm}
\end{figure}
We consider two possibilities, $J_w = J'_w$ and $J_w \neq J'_w$. 
When $J_w = J'_w$ we have what is commonly referred to as 
\emph{frustrated ring model}~\cite{Knysh_PRA2020, Cote_2023, balducci2024fighting,wang2025exponentialquadraticoptimalcontrol}. 
For this case we choose $J_w/J=0.5$, $J_f/J=-0.45$, as in Ref.~\cite{Cote_2023}. This choice of couplings satisfies conditions 
($0<J_f<J_w<J$ and $JJ_f>J_w^2$) which lead to the spin-glass bottleneck regime discussed in Ref.~\cite{Knysh_PRA2020}.
This model also exhibits a {\em reflection symmetry} around the dashed line in Fig.~\ref{fig:ring}. 
When $J_w \neq J'_w$, the reflection symmetry is broken.  
In particular, in this case, we choose $J_w'/J=0.55$. 
If we consider the spectral gap between the instantaneous ground and first excited state as a function of $s$ (see Eq.~\eqref{eqn:isingmodelham}), regardless of whether the reflection symmetry is broken or not, we see two distinct small gaps, at $s_c\approx 0.5$ and at $s_b\approx 0.9$, both decreasing to $0$ when $\Nsites\to \infty$. 
Figure \ref{fig:gap_N101_unsymm} shows the two gaps for the broken symmetry case with $J_w'\neq J_w$ for $\Nsites = 101$.
The corresponding result for $J_w'=J_w$ was given in Fig.~2 of Ref.~\cite{wang2025exponentialquadraticoptimalcontrol}. 
\begin{figure}
\centering
\includegraphics[width=\columnwidth]{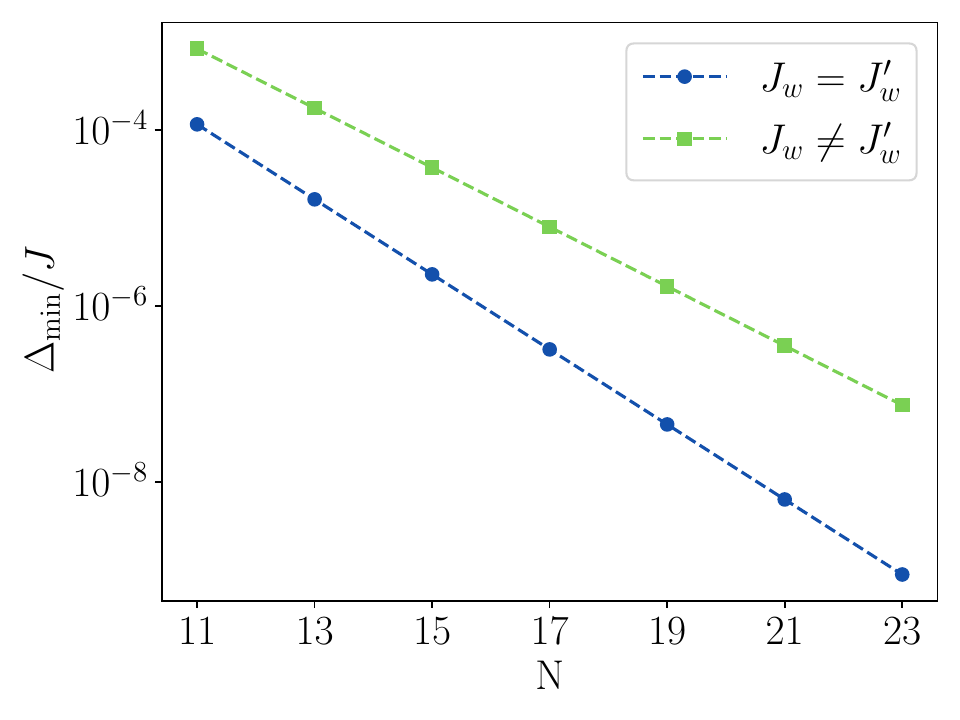}
\caption{Minimum gap $\Delta_{\min}$ as a function of $\Nsites$ for the frustrated Ising chain with and without reflection symmetry. In both cases, the small gap gets exponentially small as $\Nsites$ increases. In both models we put $J_w/J = 0.5$ and $J_f/J = 0.45$. For the chain without reflection symmetry we take $J'_w/J = 0.55$.}
\label{fig:gaps_uniform_model}
\end{figure}

The first small gap, scaling as $1/\Nsites$, marks the transition from the quantum paramagnetic to the ferromagnetic phase and is located at $s_c\approx 0.5$.
The second gap, which we indicate as $\Delta_{\min}$, corresponds to the actual bottleneck \cite{Knysh_PRA2020, Cote_2023, wang2025exponentialquadraticoptimalcontrol}, and decreases exponentially when $\Nsites\to \infty$. 
Figure~\ref{fig:gaps_uniform_model} shows that breaking the spatial symmetry with $J_w'\neq J_w$ does not qualitatively affect the spectral gap, in particular the bottleneck $\Delta_{\min}$ decreases exponentially to 0 as $\Nsites\to \infty$. 


\begin{figure}[!htp]
\centering
\begin{tikzpicture}[scale=1.7,line cap=round,line width=2pt]
\filldraw [fill=black!0!white] (0,0) circle (2cm);
\draw [line width = 0.3mm, draw=blue, dashed] (-2.2,0) -- (2.0,0) node[right, black] {};
\foreach \x [evaluate=\x as \angle using (\x-0.5)*360/13] in {1,...,13}
{
\draw[line width=1pt,fill=white] ({2*cos(\angle)},{2*sin(\angle)}) circle (3mm);
}
\node[draw=none,font=\tiny,text=black,scale=1.5] at ({2*cos(0.5*360/13)},{2*sin(0.5*360/13)}) {$1$};
\node[draw=none,font=\tiny,text=black,scale=1.5] at ({2*cos(1.5*360/13)},{2*sin(1.5*360/13)}) {$2$};
\node[draw=none,font=\tiny,text=black,scale=1.5] at ({2*cos(2.5*360/13)},{2*sin(2.5*360/13)}) {$\cdots$};
\node[draw=none,font=\tiny,text=black,scale=1.5] at ({2*cos(4.5*360/13)},{2*sin(4.5*360/13)}) {$\cdots$};
\node[draw=none,font=\tiny,text=black,scale=1.5] at ({2*cos(5.5*360/13)},{2*sin(5.5*360/13)}) {$\scriptstyle{\frac{\Nsites-1}{2}}$};
\node[draw=none,font=\tiny,text=black,scale=1.5] at ({2*cos(6.5*360/13)},{2*sin(6.5*360/13)}) {$\scriptstyle{\frac{\Nsites+1}{2}}$};
\node[draw=none,font=\tiny,text=black,scale=1.5] at ({2*cos(7.5*360/13)},{2*sin(7.5*360/13)}) {$\scriptstyle{\frac{\Nsites+3}{2}}$};
\node[draw=none,font=\tiny,text=black,scale=1.5] at ({2*cos(8.5*360/13)},{2*sin(8.5*360/13)}) {$\cdots$};
\node[draw=none,font=\tiny,text=black,scale=1.5] at ({2*cos(-0.5*360/13)},{2*sin(-0.5*360/13)}) {$\Nsites$};
\node[draw=none,font=\tiny,text=black,scale=1.5] at ({2*cos(-1.5*360/13)},{2*sin(-1.5*360/13)}) {$\scriptstyle{\Nsites-1}$};
\node[draw=none,font=\tiny,text=black,scale=1.5] at ({2*cos(-2.5*360/13)},{2*sin(-2.5*360/13)}) {$\cdots$};
\node[draw=none,font=\tiny,text=black,scale=1.5] at ({2.25*cos(5*360/13)},{2.25*sin(5*360/13)}) {$J$};
\node[draw=none,font=\tiny,text=black,scale=1.5] at ({2.25*cos(6*360/13)},{2.25*sin(6*360/13)}) {$J_w'$};
\node[draw=none,font=\tiny,text=black,scale=1.5] at ({2.25*cos(7*360/13)},{2.25*sin(7*360/13)}) {$J_w$};
\node[draw=none,font=\tiny,text=black,scale=1.5] at ({2.25*cos(8*360/13)},{2.25*sin(8*360/13)}) {$J$};
\node[draw=none,font=\tiny,text=black,scale=1.5] at ({2.3*cos(3*360/13)},{2.3*sin(3*360/13)}) {$\tilde{J}_{\scriptscriptstyle{\mathrm{N_{rand}}}}$};
\node[draw=none,font=\tiny,text=black,scale=1.5] at ({2.25*cos(2*360/13)},{2.25*sin(2*360/13)}) {$\tilde{J}_2$};
\node[draw=none,font=\tiny,text=black,scale=1.5] at ({2.25*cos(1*360/13)},{2.25*sin(1*360/13)}) {$\tilde{J}_1$};
\node[draw=none,font=\tiny,text=black,scale=1.5] at ({2.25*cos(0)},{2.25*sin(0)}) {$\scriptstyle{-J_f}$};
\node[draw=none,font=\tiny,text=black,scale=1.5] at ({2.3*cos(-1*360/13)},{2.3*sin(-1*360/13)}) {$\tilde{J}_{\scriptscriptstyle{\Nsites-1}}$};
\node[draw=none,font=\tiny,text=black,scale=1.5] at ({2.3*cos(-2*360/13)},{2.3*sin(-2*360/13)}) {$\tilde{J}_{\scriptscriptstyle{\Nsites-2}}$};
\node[draw=none,font=\tiny,text=black,scale=1.5] at ({2.3*cos(-2.8*360/13)},{2.3*sin(-2.8*360/13)}) {$\tilde{J}_{\scriptscriptstyle{\Nsites-\mathrm{N_{rand}}}}$};
\end{tikzpicture}
\caption{The generalized frustrated ring model with random couplings and an odd number of sites $\Nsites$. The model is reflection symmetric around the axis indicated by the dashed blue line if $J_w'=J_w$ and $\tilde{J}_j = \tilde{J}_{\Nsites-j}$}
\label{fig:ring_gen}
\end{figure}
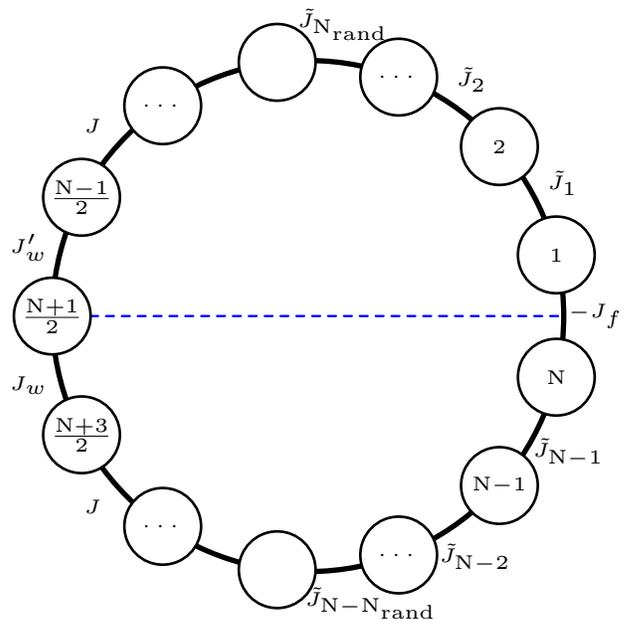

\subsubsection{Random case}
\label{subsubsec:Randomcase}

\begin{figure}
\includegraphics[width=\columnwidth]{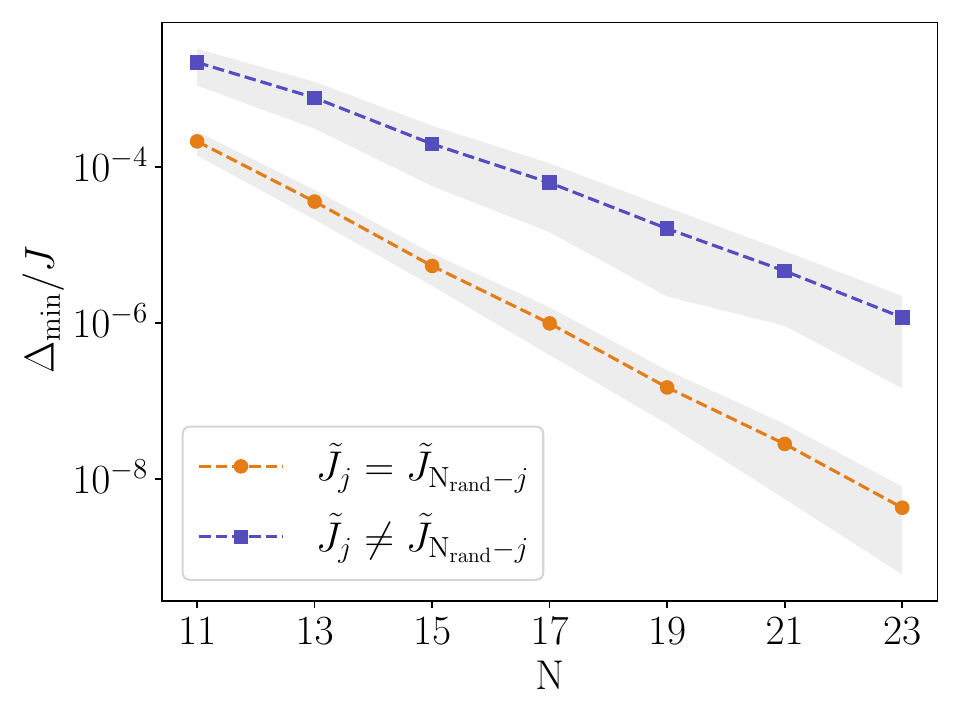}
\caption{Plot of the minimum gap $\Delta_{\min}$ as a function of $\Nsites$ for disordered frustrated Ising chains with and without reflection symmetry. In all cases, the smallest gap becomes exponentially small as $\Nsites$ increases. 
We show here 10 different random couplings configurations: the dashed lines represent the mean $\Delta_{\min}$ and the gray shadow area the standard deviation around the mean.}
\label{fig:disordered_models_gap}
\end{figure}

Next, we consider the possibility of disorder in the couplings $J_j$. 
Starting from Eq.~\eqref{eq:couplings}, we replace the couplings $J$ adjacent to the frustrated link, see Fig.~\ref{fig:ring_gen}, with random couplings $\tilde{J}_j$ and $\tilde{J}_{\Nsites-j}$, for $j = 1 \dots \Nsites_{\text{rand}} = \lfloor \frac{\Nsites-1}{4} \rfloor$ (here $\lfloor n \rfloor$ indicates the nearest smaller integer). 
These random couplings are sampled uniformly in the interval $[0.8,1]$.
If we want to impose the reflection symmetry, we set
$\tilde{J}_{\Nsites-j}=\tilde{J}_j$. 
To preserve the exponentially small gap, we leave the frustrated bond $J_{\Nsites}=-J_f$ and the weak couplings $J_w$ and $J_w'$ unchanged. 
Here we refer to these disordered models as \emph{disordered frustrated rings}.  
Even in this disordered case, the gap $\Delta_{\text{min}}$ decreases exponentially with $\Nsites$. 
Fig.~\ref{fig:disordered_models_gap} shows the average (with dashed lines, while the gray shadow represents the standard deviation) smallest spectral gaps $\Delta_{\min}$ for the symmetric and general coupling case. 

\subsection{The role of reflection symmetry}
\label{sec:reflection_symmetry}
Now we proceed to analyze the impact of reflection symmetry, $J_j=J_{\Nsites-j}$, on the computation of the critical depth. 
These calculations are general and are not limited to the frustrated models discussed in the previous section: They are valid for any quantum Ising chain model with reflection-symmetric couplings.
The reflection symmetry implies, see Appendix \ref{app:Jsymmetric_proof} for details, that the Nambu Hamiltonian $\mathH$ commutes with 
\begin{equation} \label{eqn:Pdef_1}
\mathP = \left( \begin{array}{cc} 
\mathbf{P} & \mathbf{0}_\Nsites \\
\mathbf{0}_\Nsites & \mathbf{-P} 
\end{array}
\right)\;, 
\end{equation}
where $\mathbf{P}$ is the $\Nsites \times \Nsites$ matrix with 1 in the anti-diagonal:
\begin{equation} \label{eqn:Pdef_2}
\mathbf{P} = 
 \left(
\begin{matrix}
0      & \cdots & 1      \\
\vdots & \reflectbox{$\ddots$} & \vdots \\
1      & \cdots & 0    
\end{matrix}
 \right) \;, 
\end{equation}
with $\mathbf{P}\mathbf{P} = \mathbf{1}$, and $\mathbf{P} = \mathbf{P}^{-1} = \mathbf{P}^{\dagger}$.

The fact that $[\mathH,\mathP]=0$ implies that $\mathP$ also commutes with the evolution operator $\mathU$, giving rise to further conditions beyond those implied by particle-hole symmetry. A detailed derivation,
presented in Appendix~\ref{app:Jsymmetric_proof}, shows that the dimension of the fermionic Gaussian space in presence of the parity symmetry is:
\begin{equation} \label{eqn:Pcr_symm}
\dim_{F}^{\text{symm}}  =
    \begin{cases} \displaystyle
    \frac{\Nsites^2 -1}{2}  \ 
         \ &\text{if} \ \Nsites \ \text{odd} \vspace{4mm} \\
\displaystyle         \frac{\Nsites^2}{2}
        &\text{if} \ \Nsites \ \text{even} 
    \end{cases} \;.
\end{equation}
\begin{figure*}[!ht]
\includegraphics[width=0.85\columnwidth]{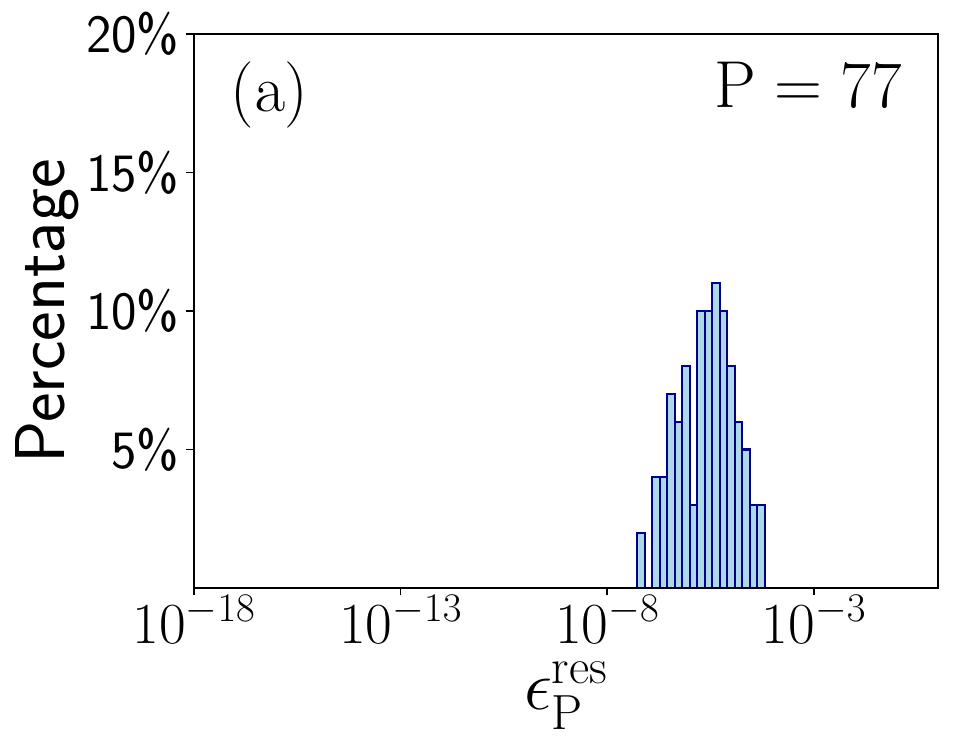}  
\includegraphics[width=0.85\columnwidth]{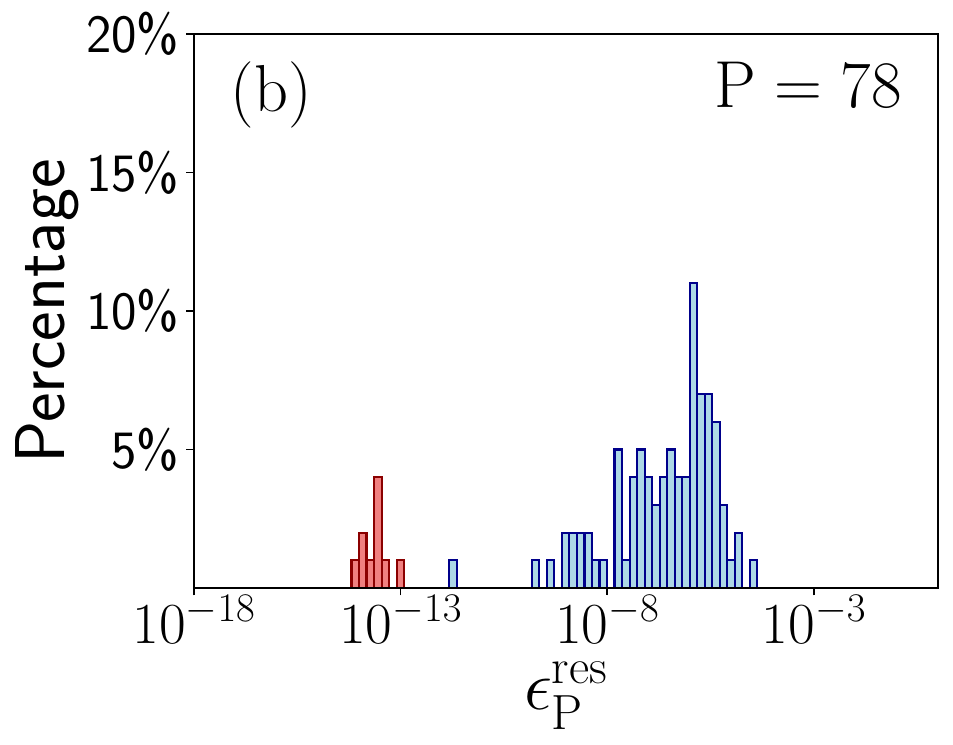} \\
\includegraphics[width=0.85\columnwidth]{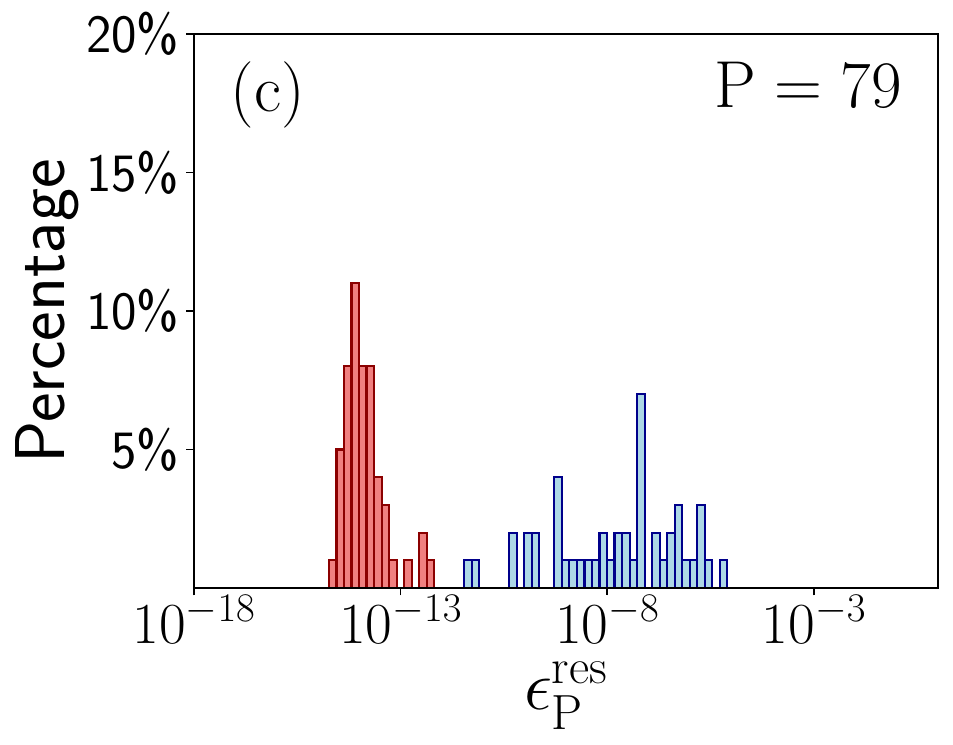}
\includegraphics[width=0.85\columnwidth]{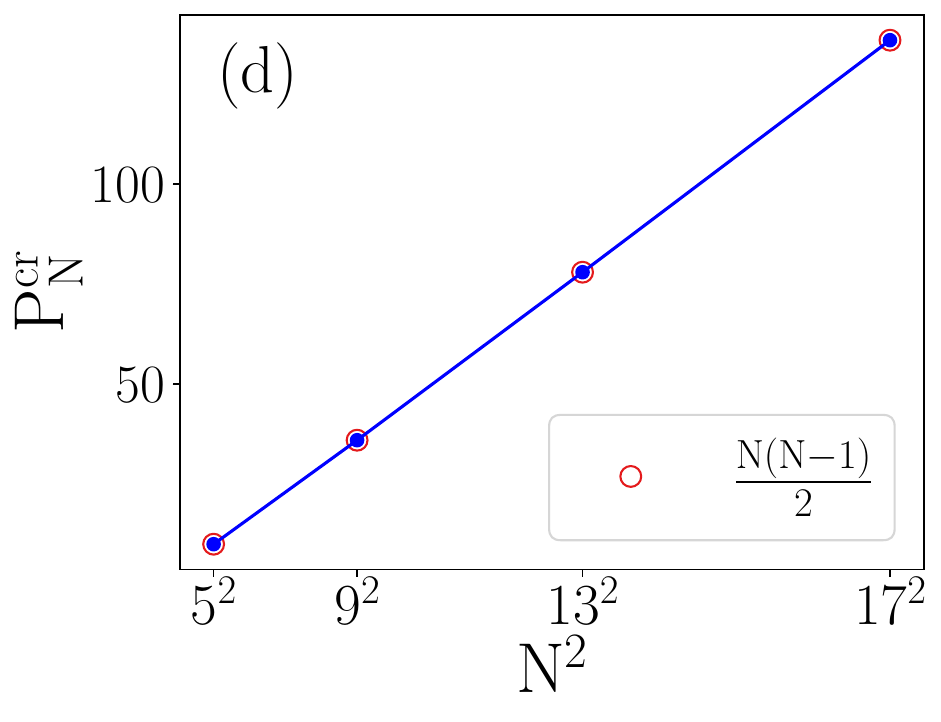}
\caption{(a)-(c) Residual energy distribution in the vicinity of $\Ptrot^{\mathrm{cr}}_{\Nsites}$ for a non-symmetric frustrated Ising ring with $\Nsites=13$ sites. Here $J_w/J = 0.5$, $J_f/J = 0.45$ and $J'_w/J = 0.55$. The distributions are obtained from $N_\text{samples} = 100$ QAOA optimizations, starting from random initial points. 
Red columns correspond to residual energies lower than or equal to $10^{-12}$, which we take as our ``numerical zero''. 
(d) Scaling of $\Ptrot^{\mathrm{cr}}_{\Nsites}$ as a function of $\Nsites$. The numerical data (solid symbols) align with the analytical expectations (empty circles) given by Eq.~\eqref{eqn:Pcr}. Here, we target the classical ferromagnetic state, i.e., $\starget=1$.}
\label{fig:P_critical_frustrated_BS}%
\end{figure*}

\begin{figure}
\includegraphics[width=\columnwidth]{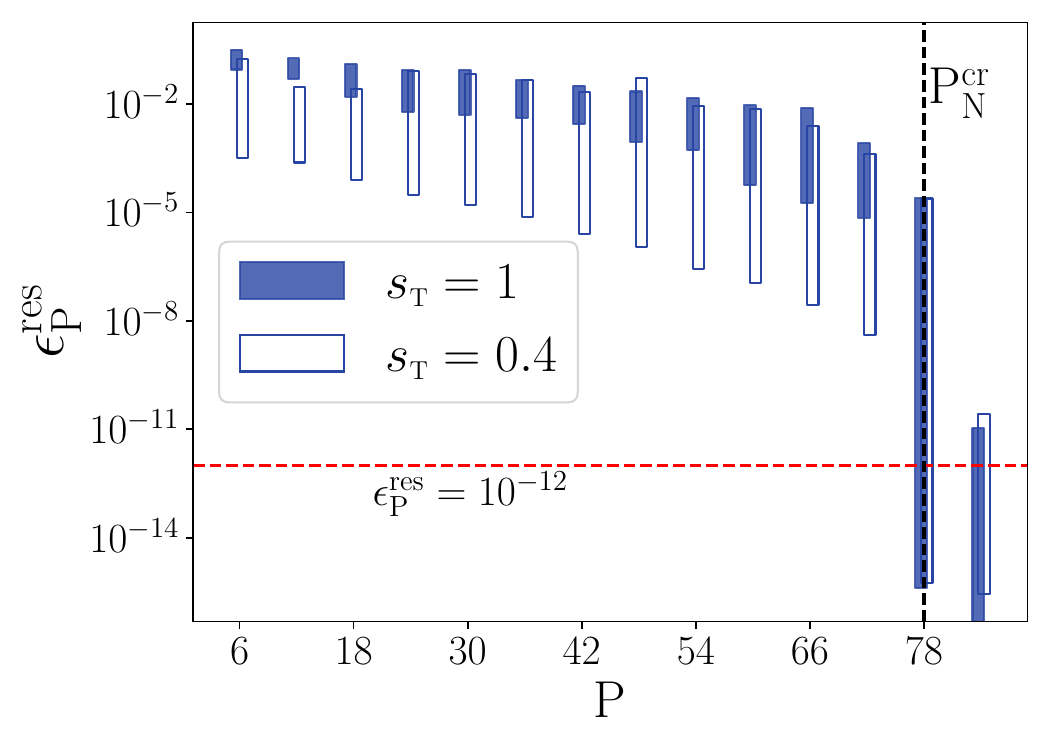}
\caption{Residual energy as a function of $\Ptrot$ for non-symmetric frustrated Ising ring with $\Nsites=13$, when targeting the ground state of $\Ho(\starget)$ with $\starget = 1$ and $\starget = 0.4$. 
Each bar indicates the range of residual energies obtained from $N_\text{samples} = 100$ QAOA optimizations, starting from random initial points. The couplings are $J_w/J = 0.5$, $J_f/J = 0.45$ and $J'_w/J = 0.55$. The value of $\Ptrot^{\mathrm{cr}}_{\Nsites}$ is same for both target states.}
\label{fig:residual_energy_comparison}
\end{figure}

Therefore we conjecture that, in the presence of reflection symmetry, we need a critical depth of a QAOA circuit to achieve controllability of the target state given by 
\begin{equation}
\Ptrot^{\mathrm{cr},\mathrm{symm}}_{\Nsites} =\frac{1}{2} \dim_{F}^{\text{symm}} \;.
\end{equation}
In the next sections, we will ``numerically prove'' these conjectures for several frustrated Ising chain models with and without randomness.
It is remarkable that 
\[ 
\Ptrot^{\mathrm{cr},\mathrm{symm}}_{\Nsites}< \Ptrot^{\mathrm{cr}}_{\Nsites} = \frac{\Nsites(\Nsites-1)}{2} \;,
\]
while the minimum gap $\Delta_{\min}$ for the symmetric model is indeed quantitatively {\em smaller} than that of the corresponding model without reflection symmetry, see Fig.~\ref{fig:gaps_uniform_model}: \emph{exact} controllability has nothing to do, evidently, with the spectral gap. 

\section{Numerical results}
\label{sec:numerical}

Our numerical validation works as follows.
We simulate the QAOA algorithm starting from random initialization of the $2\Ptrot$ parameters $\btheta$ in the range $[0,2\pi]$, then performing numerical optimization and evaluating the rescaled residual energy for the optimal parameters $\btheta^*$ obtained:
\begin{equation} \label{eqn:res_energy}
\epsilon^{\res}_{\Ptrot} (\starget)
= \frac{E_{\Ptrot}(\btheta^*) - E_{\gs}(\starget)}{\Nsites} \;,
\end{equation}
where $E_{\gs}(\starget)$ is the ground state energy of the target Hamiltonian $\Ho(\starget)$. 
We repeat this procedure for $N_\text{samples}$ different initialization of the parameters $\btheta$, building a distribution of the resulting residual energies. 
We {\em empirically} define the critical depth $\Ptrot^{\mathrm{cr}}_{\Nsites}$ as the minimum depth at which at least one out of the $N_\text{samples}$ initializations yields optimized parameters corresponding to a residual energy below $10^{-12}$, which we set as our ``numerical zero''.  
It is important to remark that since the definition of numerical zero we are adopting is empirical it is possible that for $s_{\target} \ll 1$ residual energies that are lower than $10^{-12}$ could be obtained also for $\Ptrot < \Ptrot^{\mathrm{cr}}_{\Nsites}$. Indeed, as $s_{\target}$ decreases the difference between target ground state energy and $\Ho_x$ ground state energy becomes negligible.
The value of $N_\text{samples}$ is chosen such that a further increase would not affect our result: Practically, $N_\text{samples}=100$ is enough. 

For the \emph{symmetric frustrated ring}, the numerical results that  benchmark the analytical prediction reported in Eq.~\eqref{eqn:Pcr_symm} have already been shown in Ref.~\cite{wang2025exponentialquadraticoptimalcontrol}. 
We therefore start our analysis from the frustrated ring
model {\em without} reflection symmetry, to validate
the analytical prediction in Eq.~\eqref{eqn:Pcr}.

\subsection{Frustrated ring without reflection symmetry}
We consider here the frustrated ring model with $J_w'\neq J_w$, where reflection symmetry is broken. 

\subsubsection{Targeting the classical ferromagnet: $\starget=1$.}
We start fixing $\starget = 1$, hence we target the classical ferromagnetic ground state of the Hamiltonian is $\Ho(\starget) = \Ho_{z}$.
In Fig.~\ref{fig:P_critical_frustrated_BS} (a)-(c) we show the distribution of the residual energies computed over a sample of $N_\text{samples} = 100$  QAOA optimizations from random initial points in the range 
$[0,2\pi]$ for different circuit depths $\Ptrot$. 
Figure~\ref{fig:P_critical_frustrated_BS}(b) shows that a subset of initializations leads, after local optimization of the QAOA parameters, to ``numerically zero'' residual energy when the depth $\Ptrot$ of the associated quantum circuit meets the critical value 
$\Ptrot^{\mathrm{cr}}_{\Nsites}=\frac{\Nsites(\Nsites-1)}{2}$,
see Eq.~\eqref{eqn:Pcr}, hence $\Ptrot^{\mathrm{cr}}_{\Nsites}=78$ for $\Nsites=13$. 
This fraction increases as $\Ptrot$ grows, as shown in
Fig.~\ref{fig:P_critical_frustrated_BS}(c).
Fig.~\ref{fig:P_critical_frustrated_BS}(a) shows that
no sample reaches the ground state when $\Ptrot<\Ptrot^{\mathrm{cr}}_{\Nsites}$.
In panel (d) of the same figure, we show the scaling of
$\Ptrot^{\mathrm{cr}}_{\Nsites}$ as a function of $\Nsites^2$. For each $\Nsites$, the numerical results perfectly align with the analytical prediction, indicated by empty circles in the plot. 

\subsubsection{Quantum state preparation: $\starget\neq 1$}
In this section, we show that our results hold for generic values of $s$ in $\Ho(s)$, when the target state is not classical. 
In Fig.~\ref{fig:residual_energy_comparison}, we show the residual energies for $\starget = 1$ compared to those for $\starget = 0.4$. As we expect, optimization yields better residual energies for $\starget=0.4$ at a given circuit depth $\Ptrot$ due to two main reasons: the fact that the target state for $\starget = 0.4$ is closer to the ground state of $\Ho_x$ and the fact that the spectrum of the annealing Hamiltonian does not present closing gaps between $s = 0$ and $\starget = 0.4$. However, the $\Ptrot^{\mathrm{cr}}_{\Nsites}$ required to reach the ``numerically zero'' residual energy is the same for for both the target states. This result emphasizes the fundamental difference between QAOA and analog quantum annealing: in the digital case, the resources required to exactly reach the target state scale independently of the minimum gap of the annealing Hamiltonian driving the analog Schr\"odinger dynamics.

\subsection{Disordered frustrated ring}

\begin{figure}
\includegraphics[width=\columnwidth]{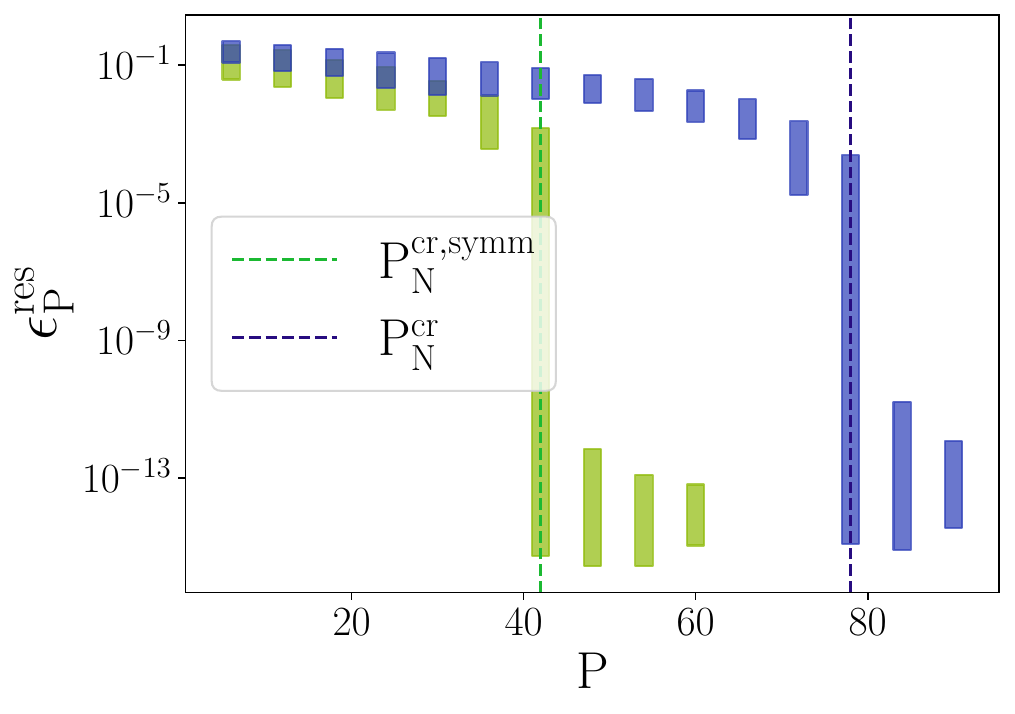}
\caption{
Residual energy as a function of $\Ptrot$ for random frustrated Ising rings with $\Nsites=13$ sites, both in the reflection symmetric (green bars) and in the symmetry broken case (blue bars). We considered 10 different coupling configurations. For each coupling realization, we performed $N_\text{samples} = 100$ QAOA optimizations from random initial points. The range bars show all the final residual energies obtained. 
At the appropriate value of $\Ptrot^{\mathrm{cr}}_{\Nsites}$, given by Eq.~\eqref{eqn:Pcr_symm} for the reflection symmetric case and by Eq.~\eqref{eqn:Pcr} for the general case, the residual energy drops down under $10^{-12}$ for all the configurations.}
\label{fig:disordered_models}
\end{figure}

We now test our analytical predictions for frustrated rings with disordered couplings, as described in Sec.\ref{subsubsec:Randomcase}.
 
Even in this disordered case the analytical predictions, for both symmetric and non-symmetric cases, are confirmed. 
The numerical test consists of taking $10$ different random coupling configurations and, for each of them, performing $N_\text{samples} = 100$ QAOA optimizations from random initial points. Figure~\ref{fig:disordered_models} shows the range of all residual energies versus $\Ptrot$ when $\Nsites=13$, both imposing reflection symmetry to the couplings (green bars) as well as with general couplings (blue bars). 
In both cases, the residual energy drops below a threshold of ``numerical zero'', when $\Ptrot$ reaches
the appropriate critical value $\Ptrot^{\mathrm{cr}}_{\Nsites}$, given by Eq.~\eqref{eqn:Pcr_symm} for the symmetric reflection case and by Eq.~\eqref{eqn:Pcr} for the general case.

Once again, as observed for the model without randomness, we notice that the reflection symmetric case shows, on average, a smaller spectral gap $\Delta_{\min}$, as shown in Fig.\ref{fig:disordered_models_gap}, while having a smaller critical value $\Ptrot^{\mathrm{cr}}_{\Nsites}$.
%
 
\section{Summary, discussion and perspectives}
\label{sec:summary}

In this work we have derived an algebraic criterion to calculate the minimal number of unitaries needed to reach the ground state with QAOA protocols for Ising chain models which can be mapped onto free-fermion Hamiltonians. 
In particular, we studied three types of frustrated-ring models, proposed as the simplest class of Ising spin systems whose minimum spectral gap decreases exponentially with system size $\Nsites$. 
We found that the critical depth required $\Ptrot^{\mathrm{cr}}_\Nsites$ grows quadratically with the system size: 
$\Ptrot^{\mathrm{cr}}_\Nsites=\Nsites(\Nsites-1)/2$, 
for the general case, and $\Ptrot^{\mathrm{cr},\mathrm{symm}}_{\Nsites} =
    (\Nsites^2-1)/4$ (odd $\Nsites$) or $\Nsites^2/4$ (even $\Nsites$),
for the case where the couplings have reflection symmetry.
%
This critical value of $\Ptrot$ is not related in any way with the spectral gap of the annealing Hamiltonian, but is entirely determined by the dimension of the controllable unitary group generated by the {\em Ansatz} gates. 

We numerically confirmed that with QAOA at $\Ptrot^{\mathrm{cr}}_{\Nsites}$ the target is reached. To do that we performed numerical minimization over many random initial parameter sets $\btheta$, followed by local optimizations. 
We further showed that $\Ptrot^{\mathrm{cr}}_{\Nsites}$ is unchanged when the target state is non-classical --- i.e., is the ground state of a linear combination of mixer and target Hamiltonian ($\starget\neq1$) --- or when moderate disorder is introduced in the couplings.

These findings highlight a fundamental distinction between quantum annealing algorithms, whose efficiency is severely limited by exponentially small spectral gaps, and digital quantum algorithms like QAOA, which can bypass this limitation through sufficiently expressive ansatzes and efficient exploration of their underlying Lie algebra~\cite{Dalessandro2007,lloyd2014information}.
In particular our findings show that the critical depth saturates the theoretical lower bound predicted in Ref.~\cite{lloyd2014information} by information theory considerations. 
This bound identifies the minimum number of evolution control parameters needed to reach a quantum state with finite accuracy, based on the dimension of the reachable manifold. In the case of QAOA applied to Gaussian fermionic systems, this bound becomes explicit and computable through the algebraic dimension of the accessible unitary group.

The quadratic scaling of the critical depth $\Ptrot^{\mathrm{cr}}_{\Nsites}$ with the system size is directly due to the fact that the dynamics is confined within the Gaussian manifold, which follows from the free-fermionic nature of the Hamiltonians involved. Future perspectives include exploring whether some kind of polynomial scaling persists also in Bethe-ansatz integrable systems --- such as the Heisenberg chain --- which lie beyond the free-fermion framework.

Also, the unitary evolution operator generated by nested commutators of the classical interaction term $\Ho_z$ and transverse field term $\Ho_x$ share the same symmetries as $\Ho_z$ and $\Ho_x$. 
This implies the possibility of similar quadratic controllability of free fermion models using digitized counter-diabatic protocols such as 
DC-QAOA~\cite{ChandaranaDigitized2022}, 
since the adiabatic gauge potentials~\cite{ClaeysFloquet2019} are defined as such nested commutators.

Finally, it will be interesting to reconsider the class of integrable time-dependent models considered in Refs.~\cite{Sinitsyn_PRA2014,Sinitsyn_PRL2018,Sinitsyn_PRL2018b}. 
The prototypical model belonging to such a class is given by $\Ho_{\scriptscriptstyle \mathrm{BCS}} = \sum_j \epsilon_j \hat{S}^z_j - (g/t) \sum_{j\neq j'} \hat{S}^+_j\hat{S}^-_{j'}$ where $\hat{S}^{x,y,z}_j$ are spin-1/2 operators at site $j$, a spin-model which turns out to be equivalent, through Anderson's pseudo-spin mapping~\cite{Anderson_PR1958}, to the BCS theory of superconductivity, in the time-independent case. 
The peculiar coupling $g/t$,  is at the root of integrability~\cite{Sinitsyn_PRL2018}. In fact is such that the fully-connected mixing term $\Ho_{\mix}=-\sum_{j\neq j'} \hat{S}^+_j\hat{S}^-_{j'}=(\hat{S}^z)^2-\hat{\mathbf{S}}^2+\Nsites/2$ --- whose ground state $|\psi_0\rangle$ has maximum total spin $\hat{\mathbf{S}}^2=\frac{\Nsites}{2}(\frac{\Nsites}{2}+1)$ with total $\hat{S^z}=0$, hence an entangled superposition of all configurations with zero magnetization --- dominates for $t=0^+$ and is switched off to zero for $t\to \infty$. 
The crucial question, in the framework of our study, would be if and how the time-dependent integrability for such models gives rise, in the digitized setting, to a peculiar controllability dependence of the target state on the system size $\Nsites$.  

%
%

\begin{acknowledgments}
G.E.S. acknowledges financial support from PNRR MUR project PE0000023-NQSTI. G.E.S. and G.P. acknowledge financial support from PRIN 2022H77XB7 of the Italian Ministry of University and Research, and from the QuantERA II Programme STAQS project that has received funding from the European Union’s H2020 research and innovation programme under Grant Agreement No 101017733.
G.E.S. acknowledges that his research has been conducted within the framework of the Trieste Institute for Theoretical Quantum Technologies (TQT).
\end{acknowledgments}

\appendix 
\section{Heisenberg QAOA dynamics}
\label{app:Heisenberg_qaoa_dynamics}
To get Eq.~\eqref{eqn:nambu_qaoa_energy} we start from Eq.\eqref{eqn:qaoa_energy},
\begin{align}
   E_{\Ptrot}(\starget) &\equiv \braket{\psi_{\Ptrot}^{\QAOA}(\btheta) | \Ho(\starget) | \psi_{\Ptrot}^{\QAOA}(\btheta)} \nonumber \\
   &= \braket{\psi_0| \Uo_{\text{ev}}^{\dagger}(\btheta) \, \Ho(\starget) \, \Uo_{\text{ev}}(\btheta) | \psi_{0}} \;,
\end{align}
where 
\[
 \Uo_{\text{ev}}(\btheta) =  \Uo_{\Ptrot}(\theta^x_\Ptrot,\theta^z_\Ptrot) \, 
\cdots \, \Uo_1(\theta^x_1,\theta^z_1) \;,
\]
is the QAOA digitized evolution operator. 
Next, we move to the Heisenberg picture for the operators. 
Considering that $\Ho(\starget)=\opbfPsidag{} \, \mathH(\starget) \, \opbfPsi{}$ is the fermionic Nambu form of the target Hamiltonian in terms of the usual $2\Nsites\times 2\Nsites$ matrix $\mathH(\starget)$, then 
\[
\Uo_{\text{ev}}^{\dagger}(\btheta) \, \Ho(\starget) \, \Uo_{\text{ev}}(\btheta)
= \sum_{j,j'=1}^{2\Nsites} \opbfPsidag{j'\Heis} \mathH_{j'j}(\starget) \opbfPsi{j\Heis} \;, 
\]
where 
\[
\opbfPsi{j\Heis}(\btheta) = 
\Uo_{\text{ev}}^{\dagger}(\btheta) \, \opbfPsi{j} \, \Uo_{\text{ev}}(\btheta) \;,
\]
is the Heisenberg form of the (Nambu) fermionic operators. 
In precise analogy with the time-dependent Bogoljubov-de Gennes (BdG) theory~\cite{mbeng2024quantum} --- viewing the digital evolution as a particular stepwise time-dependence of the Hamiltonian --- we know that 
\begin{equation}
    \opbfPsi{\Heis}(\btheta) = 
        \mathU(\btheta) \, \opbfPhi{}, 
\end{equation}
where $\opbfPhi{} =  (\opgamma{1}, \dots, \opgamma{\Nsites}, \opgammadag{1}, \dots, \opgammadag{\Nsites})^\transpose$ are the Bogoljubov fermions for which the initial state $|\psi_0\rangle$ is the vacuum, i.e., such that
$\opgamma{j}|\psi_0\rangle=0$, and
$\mathU(\btheta)$ is a $2\Nsites\times 2\Nsites$ unitary operator solving the
BdG equations:
\begin{equation} \label{eq:evolutionboldtheta}
    \mathU(\btheta) =
\mathU_\Ptrot(\theta^x_\Ptrot,\theta^z_\Ptrot) \, \dots \, 
    \mathU_1(\theta^x_1,\theta^z_1) \, \mathU_0
    \,
\end{equation}
with
\begin{equation} \label{eqn:U_nambu_app}
        \mathU_p(\theta^x_p,\theta^z_p) = \nep^{-2i \theta^x_p \mathH_x \,} \nep^{-2i \theta^z_p \mathH_z} \;.
\end{equation}
The initial unitary $\mathU_0$ is fixed by finding $\opbfPhi{}\!\!$ in terms of the original fermions $\opbfPsi{}\!\!$. 
Since $|\psi_0\rangle$ is the ground state of
\[ 
\Ho_x=-h\sum_{j=1}^{\Nsites}
(\opc{j}\opcdag{j}-\opcdag{j}\opc{j}) \;,
\]
we easily conclude that $|\psi_0\rangle=|0\rangle$ if $h>0$, while $\ket{\psi_0}=\opcdag{\Nsites}\cdots\opcdag{1}|0\rangle$ if $h<0$. Hence, the Bogoljubov transformation simply reads:
\begin{equation}
\mathU_0 = 
    \left( \begin{array}{cc} \mathbf{1}_\Nsites & \mathbf{0}_\Nsites \\
    \mathbf{0}_\Nsites & \mathbf{1}_\Nsites \end{array}   \right) \;, \hspace{0.5cm} \text{if } h > 0 \;,
\end{equation}
and
\begin{equation}
\mathU_0 = 
    \left( \begin{array}{cc} \mathbf{0}_\Nsites & \mathbf{1}_\Nsites \\
    \mathbf{1}_\Nsites & \mathbf{0}_\Nsites \end{array}   \right) \;, \hspace{0.5cm} \text{if } h < 0 \;,
\end{equation}
where $\mathbf{0}_\Nsites$ and $\mathbf{1}_\Nsites$ are, respectively, the null and the identity matrix of dimension $\Nsites\times\Nsites$.

Using these steps, we can write:
\begin{equation}
E_{\Ptrot}(\starget) =
\sum_{j,j'=1}^{2\Nsites} 
\Big( \mathU^{\dagger}(\btheta) \,  \mathH(\starget) \, \mathU(\btheta) \Big)_{jj'} 
\mathGamma_{jj'} \;,
\end{equation}
where the Green's function 
\begin{equation}
\label{eqn:gammadefinition}
    \mathGamma_{jj'} = 
    \langle \psi_0| \opbfPhidag{j} \opbfPhi{j'} | \psi_0\rangle \;,
\end{equation}
is easily calculated:
\begin{equation}
\mathGamma = \left( 
\begin{array}{cc}  \mathbf{0}_\Nsites & \mathbf{0}_\Nsites \\                        \mathbf{0}_\Nsites & \mathbf{1}_\Nsites 
\end{array}
\right) \;.
\end{equation}
Since $\mathGamma$ is symmetric, we can finally write:
\begin{equation}
    E_{\Ptrot}(\starget) =
\Trace
\Big( \mathU^{\dagger}(\btheta) \, \mathH(\starget) \, \mathU(\btheta) \,
\mathGamma \Big) \;.
\end{equation}

\section{Derivation for the reflection symmetric case}
\label{app:Jsymmetric_proof}
We start showing that $\mathH_x$ and $\mathH_z$ both commutes with $\mathP$ defined in Eqs.~\eqref{eqn:Pdef_1}-\eqref{eqn:Pdef_2}.
For $\mathH_x$ this is straightforward, since:
\begin{equation}
\mathP \mathH_x \mathP 
= \left( \begin{array}{cc} 
h\mathbf{P}\mathbf{P} & \mathbf{0}_\Nsites \\
\mathbf{0}_\Nsites & -h\mathbf{P}\mathbf{P} 
\end{array}
\right) 
= \left( \begin{array}{cc} 
h\mathbf{1}_{\Nsites} & \mathbf{0}_\Nsites \\
\mathbf{0}_\Nsites & -h\mathbf{1}_\Nsites 
\end{array}
\right) = \mathH_x \;. \nonumber 
\end{equation}
For $\mathH_z$ we can write:
\begin{equation}
\label{eqn:paritytransofrmationHz}
    \mathP\mathH_z\mathP = \left( \begin{array}{cc} 
\mathbf{P}\mathbf{A}_z \mathbf{P} & -\mathbf{P}\mathbf{B}_z\mathbf{P} \\
\mathbf{P}\mathbf{B}_z\mathbf{P} & -\mathbf{P}\mathbf{A}_z\mathbf{P} 
\end{array}
\right) \;.
\end{equation}
Using the definition of $\mathbf{P}$, we obtain $(\mathbf{P}\mathbf{A}_z\mathbf{P})_{jj'} = (\mathbf{A}_{z})_{ \Nsites+1-j, \Nsites+1-j'}$. It follows that:
\begin{equation} \label{eqn:Azsymmetryconditions}
\begin{aligned}
(\mathbf{P}\mathbf{A}_z\mathbf{P})_{j,j+1} &= (\mathbf{A}_z)_{\Nsites+1-j,\Nsites-j} = -J_{\Nsites-j} \equiv -J_{j}\\
(\mathbf{P}\mathbf{A}_z\mathbf{P})_{j+1,j} &= (\mathbf{A}_z)_{\Nsites-j,\Nsites+1-j} = -J_{\Nsites-j} \equiv -J_{j}\\
(\mathbf{P}\mathbf{A}_z\mathbf{P})_{\Nsites,1} &= (\mathbf{A}_z)_{1,\Nsites} = (-1)^\text{p} J_{\Nsites}/2\\
(\mathbf{P}\mathbf{A}_z\mathbf{P})_{1,\Nsites} &= (\mathbf{A}_z)_{\Nsites,1} = (-1)^\text{p} J_{\Nsites}/2 \\
\end{aligned} \;.
\end{equation}
Comparing Eq.~\eqref{eqn:Azsymmetryconditions}, with  Eqs.~\eqref{eqn:AB_z} and \eqref{eqn:A_zborder} we find that $\mathbf{P}\mathbf{A}_z\mathbf{P} = \mathbf{A}_z$.
For the case of $\mathbf{P}\mathbf{B}_z \mathbf{P}$, the derivation is similar, with the result $\mathbf{P}\mathbf{B}_z \mathbf{P} = -\mathbf{B}_z$. Applying these results to 
Eq.~\eqref{eqn:paritytransofrmationHz} we obtain that $\mathP \mathH_z\mathP = \mathH_z$.

Now we need to distinguish the case in which $h>0$, and $\mathU_0 = \mathbb{1}$, from the case in which $h <0$ where
\begin{equation} \label{eqn:U0_h_negative}
\mathU_0 = 
\left( 
\begin{array}{cc} \mathbf{0} & \mathbf{1} \\
\mathbf{1} & \mathbf{0} 
\end{array}\right) \;. 
\end{equation} 
We start by considering $h > 0$, where 
$\left[ \mathU_0, \mathP\right]=0$. 
Since $\mathH_x$, $\mathH_z$ and $\mathU_0$ all commute with $\mathP$, and $\mathU$ is constructed from a product of $\mathU_0$ and of exponentials of $\mathH_z$ and $\mathH_x$, it follows that 
\begin{equation} \label{eqn:ref_symmetry_requirement}
    [\mathU(\btheta), \mathP] = 0 \;.
\end{equation} 

This implies that to calculate the dimension of the group we will have to consider both the particle-hole symmetry and the reflection symmetry induced by
$[\mathU(\btheta), \mathP]=0$.

If we consider a generic 
$\mathU =  \left( 
\begin{array}{cc}  
\mathbf{U} & \mathbf{V^*} \\
\mathbf{V} & \mathbf{U^*}
\end{array}
\right)$ 
and impose that
$\left[\mathU, \mathP\right] = 0$, 
we will find the following conditions:
\begin{equation}
    \left( 
\begin{array}{cc}  
\mathbf{UP -PU} & \mathbf{-V^*P - PV^*} \\
\mathbf{VP + PV} & \mathbf{-U^*P + PU^*} 
\end{array}
\right) = \left( 
\begin{array}{cc}  
\mathbf{0}_\Nsites & \mathbf{0}_\Nsites \\
\mathbf{0}_\Nsites & \mathbf{0}_\Nsites 
\end{array}
\right) \;,
\end{equation}
which imply that $\left[\mathbf{U}, \mathbf{P}\right] = 0$ and $\{\mathbf{V}, \mathbf{P}\} = 0$. 
Therefore, due to this additional symmetry of the group, the number of constraints to calculate the dimension of the group increases:
\begin{equation}
\label{eqn:conditionsymmetry}
    \left\{
    \begin{aligned}
        \U^{\dagger}\U + \V^{\dagger}\V &= \mathbf{1}\\
        \U^{\dagger}\V^* + \V^\dagger\U^* &= \mathbf{0}\\
        [\U, \mathbf{P}] &= \mathbf{0}\\
        \{\V, \mathbf{P}\} &= \mathbf{0}
    \end{aligned}
    \right. \;.
\end{equation}
The first two systems of constraints are due to the usual p-h symmetry and unitarity,
see Eq.~\eqref{eqn:ph-symmetry-constraint}.
Now we analize the implications of the set of new constraints due to reflection symmetry, 
i.e., $[\U, \mathbf{P}] = \mathbf{0}$
and $\{\V,\mathbf{P}\} = \mathbf{0}$.

The condition $[\U, \mathbf{P}] = \mathbf{0}$ implies that $\mathbf{P}\mathbf{U}\mathbf{P} = \mathbf{U}$ and therefore, applying the definition of $\mathbf{P}$, we obtain  $\U_{jj'} = \U_{\Nsites+1-j,\Nsites+1-j'}$.
The condition 
$\{\V,\mathbf{P}\} = \mathbf{0}$ implies that $\mathbf{P}\mathbf{V}\mathbf{P = - \mathbf{V}}$ and therefore $\V_{jj'} = -\V_{\Nsites+1-j,\Nsites+1-j'}$. 
These two conditions reduce by a factor $2$ the total number of real free parameter that the two complex $\Nsites \times \Nsites$ matrices $\U$ and $\V$ can have, for both even and odd $\Nsites$. 
As a consequence, 
starting from $4 \Nsites^2$ real parameters, needed to specify $\U$ and $\V$,
by applying the reflection symmetry, we are 
imposing $2\Nsites^2$ constraints, remaining with a total number of parameters:
\begin{equation}
    \mathrm{N}_{\textrm{symm}} = 
    2\Nsites^2 \;.
\end{equation}

Now in order to see how many constraints the first two conditions in Eq.~\eqref{eqn:conditionsymmetry} are setting, we need to take into account that, having already imposed $\left[\U,\bfP\right]=0$, 
$\{\V,\bfP\}=0$, we have reduced the number of degrees of freedom of $\U$ and $\V$, and therefore also the number of independent variables of $\U^{\dagger}\U + \V^{\dagger}\V$ and $\U^{\dagger}\V^* + \V^{\dagger}\U^*$.
Since $\mathbf{P} = \bfP^{\dagger} = \bfP^*$,  it follows that $\left[\U, \bfP\right] = 0$ and $\{\V,\bfP\} = 0$ imply:
\begin{equation}
\begin{aligned}
\left[\U^*, \bfP\right] &= \left[\U^{\dagger}, \bfP\right] = 0\\
\{\V^{*},\bfP\} &= \{\V^{\dagger},\bfP\} = 0
\end{aligned} \;.
\end{equation}
 Now we set $\mathbf{K} = \U^{\dagger}\U + \V^{\dagger}\V=\bfK^{\dagger}$, an Hermitian matrix.
 We can verify that $[\mathbf{K}, \mathbf{P}] = \mathbf{0}$. This follows from the fact that
 \begin{equation}
 \begin{aligned}
     \U^{\dagger} \U \bfP &= \U^{\dagger}\bfP \U = \bfP \U^{\dagger}\U,\\
     \V^{\dagger}\V \bfP &=-\V^{\dagger}\bfP \V = \bfP \V^{\dagger}\V
 \end{aligned}
 \end{equation}
If we combine both $\left[\mathbf{P}, \bfK\right] = 0$ and $\bfK^{\dagger} = \bfK $, we find:
%
%
\begin{equation}
\begin{aligned}
    \mathbf{K}_{jj'} &= \mathbf{K}^*_{j'j}\\
    \bfK_{jj'} &= \bfK^*_{\Nsites+1-j' \Nsites+1-j} 
\end{aligned} \;.
\end{equation}
Therefore, $\mathbf{K}$ is simultaneously Hermitian under reflection across the main diagonal and the antidiagonal.
Now we first consider the case in which $\Nsites$ is odd. 
To fully determine $\bfK$ we need to specify $\Nsites$ real parameters to fill the diagonal and the antidiagonal, that are both real. 
For the remaining entries in the matrix, we need to consider the total number of entries, $\Nsites^2$, remove both diagonals ($2\Nsites-1$) paying attention to the fact that they share a common entry, divide by $4$, since only one quarter needs to be specified, and then multiply by $2$, to count the number of real parameters. Therefore, the total number of independent real parameters of $\bfK$ is
$\Nsites + \frac{\Nsites^2 - 2 \Nsites +1}{2}$. The case of $\Nsites$ even is similar, but this time the number of entries to specify both diagonals is $2 \Nsites$, as they do not share an entry. 
Consequently, the number of independent parameters of $\mathbf{K}$ depends on the parity of $\Nsites$:
\begin{equation}
\mathrm{N_K} = 
\begin{cases}
    \Nsites + \frac{(\Nsites^2 -2\Nsites +1)}{2} = \frac{\Nsites^2+1}{2} \ &\text{if} \ \Nsites \ \text{odd} \vspace{3mm}\\
    \Nsites +  \frac{\Nsites^2 - 2\Nsites}{2} = \frac{\Nsites^2}{2} \ &\text{if} \ \Nsites \ \text{even} 
\end{cases} \;.
\end{equation}
Since $\bfK$ has $\mathrm{N_K}$ independent parameters, when we impose the first condition of Eq.~\eqref{eqn:conditionsymmetry}, we are actually imposing only $\mathrm{N_K}$ constraints. 

Next, we set
$\mathbf{Q}=\U^{\dagger}\V^* + \V^\dagger\U^*=\mathbf{Q}^\transpose$, 
a complex symmetric matrix.
In this case we have 
$\{\mathbf{Q},\mathbf{P}\}=\mathbf{0}$, as implied by: 
\begin{equation}
\label{eqn:composedcommutation}
    \begin{aligned}
        \U^{\dagger}\V^*\bfP &= -\U^{\dagger}\bfP\V = -\bfP \U^{\dagger}\V\\
        \V^{\dagger}\U^*\bfP &= \V^{\dagger}\bfP \U^* = -\bfP\V^{\dagger}\U^* 
    \end{aligned} \;.
\end{equation}
Combining $\Q = \Q^\transpose$ and
$\{\Q,\bfP\}=0$ we find that: 
%
\begin{equation}
    \begin{aligned}
        \mathbf{Q}_{jj'} &= \mathbf{Q}_{j'j}\\
        \mathbf{Q}_{jj'} &= - \mathbf{Q}_{\Nsites+1-j',\Nsites+1-j}
    \end{aligned} \;.
\end{equation}
Therefore, $\mathbf{Q}$ is complex symmetric with respect to the main diagonal and complex antisymmetric with respect to the antidiagonal. Consequently, the antidiagonal elements of $\mathbf{Q}$ are null. We first consider the case in which $\Nsites$ is odd. In this case, to fill the diagonal we need to specify $\Nsites-1$ real parameters, as the entry shared with the anti-diagonal must vanish. 
To fill the rest of the matrix, proceeding as we did for $\mathbf{K}$, we need 
$\frac{\Nsites^2-2\Nsites+1}{2}$ parameters. 
If $\Nsites$ is even, to fill the diagonal we need to specify $\Nsites$ real parameters. 
To fill the rest just 
$\frac{\Nsites^2 - 2\Nsites}{2}$ as for the $\bfK$ case. 
Therefore, the number of free parameter of $\mathbf{Q}$ is:
\begin{equation}
\label{eqn:freeparameterQ}
    \mathrm{N_Q} = 
    \begin{cases}
        \textstyle{\Nsites - 1} +  \frac{(\Nsites^2 -2\Nsites +1)}{2} = \frac{\Nsites^2-1}{2} \ &\text{if} \ \Nsites \ \text{odd} \vspace{3mm}\\
        \textstyle{\Nsites} + \frac{\Nsites^2 - 2\Nsites}{2} = \frac{\Nsites^2}{2} \ &\text{if} \ \Nsites \ \text{even} 
    \end{cases} \;.
\end{equation}
After applying the constraints imposed by reflection symmetry and by ph-symmetry plus unitarity (through $\mathbf{K}$ and $\mathbf{Q}$), the dimension of the space reads:
\begin{equation}
    \dim_U = 4\Nsites^2 - \mathrm{N}_{\textrm{symm}} - \mathrm{N_K} - \mathrm{N_Q} = \Nsites^2 \;,
\end{equation}
independent of the parity of $\Nsites$.

Now we want to calculate the dimension $\dim_W$ of the gauge-freedom matrix $\mathbb{W}$, see \eqref{eqn:W}, for $\Nsites$ even and odd. 
Reflection symmetry implies that  $[\mathW,\mathP]=0$, hence:
\begin{equation}
\label{eqn:conditionsymmetrygauge}
    \left\{
    \begin{aligned}
        \mathbf{W}^{\dagger}\mathbf{W} &= \mathbf{1}\\
        [\mathbf{W}, \mathbf{P}] &= \mathbf{0}
    \end{aligned}
    \right. \;.
\end{equation}
From the second condition in Eq.~\eqref{eqn:conditionsymmetrygauge}, we can infer that $\mathbf{W}_{jj'} = \mathbf{W}_{\Nsites+1-j, \Nsites+1-j'}$. 
Therefore, there are 
$2\times\frac{\Nsites^2+1}{2}$ free real parameters if $\Nsites$ is odd, and 
$2\times\frac{\Nsites^2}{2}$ if $\Nsites$ is even. 
Now we need to remove the parameters constrained by the first equation in Eq.~\eqref{eqn:conditionsymmetrygauge}.
The matrix $\mathbf{W}^{\dagger}\mathbf{W}$ has the same symmetries as $\mathbf{K}$. 
Therefore:
\begin{equation}
    \dim_W = 
    \begin{cases}
        \Nsites^2 +1 - \mathrm{N_K} = \frac{\Nsites^2+1}{2} \ &\text{if} \ \Nsites \ \text{odd} \vspace{3mm} \\
        \Nsites^2 - \mathrm{N_K}  = \frac{\Nsites^2}{2} \ &\text{if} \ \Nsites \ \text{even} 
    \end{cases} \;.
\end{equation}
Finally, we arrived at: 
\begin{equation} \label{eq:dim_Ug_frustrated}
    \dim_U - \dim_W = 
    \begin{cases}
    \frac{\Nsites^2 -1}{2}  \ 
         \ &\text{if} \ \Nsites \ \text{odd} \vspace{3mm} \\
         \frac{\Nsites^2}{2}
        &\text{if} \ \Nsites \ \text{even} 
    \end{cases} \;.
\end{equation}
As a consequence, the number of layers in the QAOA circuit is predicted to be:
\begin{equation} \label{eqn:Pcr_symm_app}
\Ptrot^{\mathrm{cr},\mathrm{symm}}_{\Nsites} 
= \frac{\dim_{U} - \dim_{W}}{2} = 
    \begin{cases} \displaystyle
    \frac{\Nsites^2 -1}{4}  \ 
         \ &\text{if} \ \Nsites \ \text{odd} \vspace{3mm} \\
\displaystyle         \frac{\Nsites^2}{4}
        &\text{if} \ \Nsites \ \text{even} 
    \end{cases} \;.
\end{equation}

The case in which $h<0$, is different from the previous, because in this case $\mathU_0$ is
given by Eq.~\eqref{eqn:U0_h_negative} and therefore $\mathP$ {\em anticommutes} with $\mathU_0$, and also with  $\mathU(\btheta)$:
\[
\{\mathU(\btheta),\mathP\}=0 \;.
\]
%
%
Imposing $\{\mathU, \mathP \} = 0$ to a 
generic $\mathU =  \left( 
\begin{array}{cc}  
\mathbf{U} & \mathbf{V^*} \\
\mathbf{V} & \mathbf{U^*}
\end{array}
\right)$, 
we obtain that $\left[\V,\bfP\right]=0$ and 
$\{\U,\bfP\}=0$. 
These opposite commutation rules, however, do not change the results in terms of dimension of the space and number of free parameters. 
Indeed, it still holds that $\left[\bfK,\bfP\right]=0$, and 
$\{\Q, \bfP\}=0$. 
Therefore, the expression for $\Ptrot^{\mathrm{cr},\mathrm{symm}}_{\Nsites}$ 
in Eq.~\eqref{eqn:Pcr_symm_app} remains the same.

Even in this symmetric case it is possible to calculate the dimension of the submanifold of the fermionic Gaussian states that respect the symmetry. Indeed we need to calculate the number of parameters of $\mathbf{Z}$, that is defined in Eq.~\eqref{eqn:gaussianstate}. As derived in Ref.~\cite{mbeng2024quantum} we can write 
\begin{equation}
\label{eq:thoulessformula}
\mathbf{Z} = \mathbf{V}^{*} \big( \mathbf{U}^{*}\big)^{-1}
 \;,
\end{equation}
where, see Eq.~\eqref{eqn:U_nambu_UV}, $\mathbf{U}$ and $\mathbf{V}$ are the $\Nsites \times \Nsites$ submatrices of
\[
\mathbb{U} =
\left(
\begin{array}{cc}   
    \mathbf{\mathbf{U}} & \mathbf{V}^* \\
\mathbf{V} & \mathbf{U}^*
\end{array} 
\right) \;.
\]

 It is possible therefore to apply the symmetry constraints to $\mathbf{Z}$. If $\mathbf{U}^{\dagger}$ commutes or anti-commutes with $\mathbf{P}$, also $\left(\mathbf{U}^{\dagger}\right)^{-1}$ respects the same commutation or anti-commutation rule. At the same time we know that  if $\mathbf{U}$ commutes with $\mathbf{P}$, $\mathbf{V}^{\dagger}$ anti-commutes with it  and viceversa. Therefore, we can infer, using the same argument as in Eq.~\eqref{eqn:composedcommutation}, that 
\begin{equation}
\label{eqn:anticommutationZ}
    \{\mathbf{Z}, \mathbf{P}\} = 0.
\end{equation}
Since $\mathbf{Z}$ is a complex antisymmetric matrix and anticommute with $\mathbf{P}$, this implies that
\begin{equation}
    \begin{aligned}
    \mathbf{Z}_{jj'} &= -\mathbf{Z}_{j'j}\\
        \mathbf{Z}_{jj'} &= \mathbf{Z}_{\Nsites+1-j',\Nsites+1-j}
    \end{aligned} \;,
\end{equation}
therefore $\mathbf{Z}$ is complex antisymmetric with respect the main diagonal and complex symmetric with respect the anti-diagonal. From that it follows that $\mathbf{Z}$ has the same number of free parameters as $\mathbf{Q}$ (see Eq.~\eqref{eqn:freeparameterQ}). We can conclude that the dimension of the fermionic Gaussian manifold coincides as expected with $\dim_{U} - \dim_{W}$:
\begin{equation}
\label{eqn:freeparameterQ}
    \dim_F  = 
    \begin{cases}
        \textstyle{\Nsites - 1} +  \frac{(\Nsites^2 -2\Nsites +1)}{2} = \frac{\Nsites^2-1}{2} \ &\text{if} \ \Nsites \ \text{odd} \vspace{3mm}\\
        \textstyle{\Nsites} + \frac{\Nsites^2 - 2\Nsites}{2} = \frac{\Nsites^2}{2} \ &\text{if} \ \Nsites \ \text{even} 
    \end{cases} \;.
\end{equation}
\section{Uniform quantum Ising chain}
We study here the case in which all the couplings are uniform, $J_j = J > 0$ for 
$j=1\cdots \Nsites$, and the chain has periodic boundary conditions.
The QAOA-controllability for this case was
already considered in Ref.~\cite{mbeng2019optimal}, but we want to re-derive the result with the techniques used here for the general case.

In the translationally invariant case, we perform a Fourier transformation of the fermionic operators in Eq.~\eqref{quadratic-H:eqn}, directly implementing the translational invariance symmetry~\cite{mbeng2024quantum}.
We assume for simplicity that $\Nsites$ is even, therefore, we transform:
\begin{equation}
\label{eqn:fouriertransform}
\opc{n} = \frac{\nep^{-i \pi/ 4}}{\sqrt{\Nsites}} \sum_{k} \nep^{ikn} \opc{k}  
\;,
\end{equation}
with $\Nsites$ wavevectors $k$ consistent with the boundary conditions. 
Hence, for even fermion parity~\cite{mbeng2024quantum}:
\begin{equation}
k \in \left(\pm \textstyle{\frac{\pi}{\Nsites}, \pm \frac{3\pi}{\Nsites},\cdots,\pm \frac{(\Nsites-1)\pi}{\Nsites}} \right) \;,
\end{equation}
while for odd fermion-parity:
\begin{equation}
k \in \left( \textstyle{\frac{(-\Nsites+1)\pi}{\Nsites}, \dots, \pi} \right) \;.
\end{equation}
We restrict ourselves to the even fermion-parity case. 
By applying the Fourier transformation to the transverse field term $\Ho_x$, we have~\cite{mbeng2024quantum}:
%
%
\begin{equation}
\label{eqn:Hxisinguniform}
 \Ho_{x} = h \sum_{k} \Big( \opcdag{k} \opc{k}  - \opc{k} \opcdag{k}  \Big)  \;.
\end{equation}
The transformed interaction term $\Ho_z$ reads
~\cite{mbeng_quantum_2019}:
%
%
%
\begin{equation}
\label{eqn:Hzisinguniform}
\begin{aligned}
\Ho_{z} = 
-J\sum_{k} 
&\cos k\left( \opcdag{k} \opc{k} 
- \opc{k} \opcdag{k} \right) \\ 
+J\sum_{k}
&\sin k \left(\opcdag{k}\opcdag{-k} 
+ \opc{-k}\opc{k} \right)\;.
\end{aligned}
\end{equation} 
%
%
To simplify our notation, we denote by
\begin{equation}
k_1 = \textstyle{\frac{\pi}{\Nsites}} \,, 
k_2 = \textstyle{\frac{3\pi}{\Nsites}} \,,
\cdots\, 
k_{n} = \textstyle{\frac{(\Nsites-1)\pi}{\Nsites}} \;,
\end{equation}
the $n=\frac{\Nsites}{2}$ positive wavevectors relevant for the even-fermion parity case.
We can write both Eqs.~\eqref{eqn:Hxisinguniform} and~\eqref{eqn:Hzisinguniform} 
in the Nambu formalism. 
First we define the Nambu vector of fermionic operators as:
\[
 \opbfPsi{}\!\! = (
 \opc{-k_1}, \opcdag{k_1}, \dots, 
 \opc{-k_n}, \opcdag{k_n}, 
 \opcdag{-k_1},\opc{k_1}, \dots,  
 \opcdag{-k_n},\opc{k_n})^\transpose \;.
\]
Then we can write, as in Eq.~\eqref{quadratic-H:eqn}:
\begin{eqnarray} \label{eqn:quadraticisinguniform}
\Ho_{x/z} = \opbfPsidag{} \, \mathH_{x/z} \, \opbfPsi{} \;.
\end{eqnarray}
For the transverse field term we have that, 
\begin{equation}
\mathH_x = 
\left( \begin{array}{cc} \A_x & \mathbf{0} \\
        \mathbf{0} & -\A_x \end{array} 
\right) \;,
\end{equation}
where $\A_x$ is purely diagonal and reads
\begin{equation}
\A_x = h \begin{pmatrix}
1 & 0 & 0 & 0 & \cdots & 0 \\
0 & -1 & 0 & 0 &\cdots & 0 \\
0 & 0 & 1 & 0 & \cdots & 0 \\
0 & 0 & 0 & -1 & \cdots & 0 \\
\vdots & \vdots & \vdots & \vdots & \ddots & \vdots \\
0 & 0 & 0 & 0 & \cdots & -1
\end{pmatrix}\;.
\end{equation}

For the interaction part we have:
\begin{equation}
\mathH_z = 
\left( \begin{array}{cc} \A_z & \mathbf{0} \\
        \mathbf{0} & -\A_z \end{array} 
\right) \;,
\end{equation} 
where $\A_z$ is a block diagonal matrix, with $2\times2$ blocks $\text{A}^{k}_z$:
\begin{equation}
\A_z =
\begin{pmatrix}
\text{A}^{k_1}_z & 0 & \cdots & 0 \\
0 & \text{A}^{k_2}_z & \cdots & 0 \\
\vdots & \vdots & \ddots & \vdots \\
0 & 0 & \cdots & \text{A}^{k_n}_z
\end{pmatrix} \;,
\end{equation}
and each block $\text{A}^{k}_z$ is defined as:
\begin{equation}
\text{A}^{k}_z = J
\begin{pmatrix}
-\cos k & \sin k \\
\sin k &  \cos k
\end{pmatrix} \;.
\end{equation}

At this point, it is convenient to define $\mathU_0$ in a way that is $2\times2$ block diagonal, so that also $\mathU(\btheta)$ will be in a $2\times2$ block diagonal form. 
To do this, we can define the Nambu vector $\opbfPhi{}\!\!$, containing the Bogoliubov operators $\opgamma{k}$ that annihilate the initial state 
(i.e. $\opgamma{k}|\psi_0\rangle=0$) as:
\[
\opbfPhi{}\!\! = 
(\opgamma{-k_1},\opgammadag{k_1}, \dots, \opgamma{-k_n},  \opgammadag{k_n}, 
\opgammadag{-k_1},\opgamma{k_1}, \dots,\opgammadag{-k_n},  \opgamma{k_n} )^\transpose \;.
\]
Since $|\psi_0\rangle$ is the ground state of $\Ho_x$, we obtain that,
\begin{equation}
    \opbfPsi{} \equiv \mathbb{U}_0 \opbfPhi{},
\end{equation}
with $\mathU_0=\mathbb{1}$ if $h>0$, and
\begin{equation}
\mathU_0 =  \operatorname{diag} \left( 
\begin{pmatrix} 0 & 1 \\ 1 & 0 \end{pmatrix}, 
\begin{pmatrix} 0 & 1 \\ 1 & 0 \end{pmatrix}, 
\dots 
\right)   \hspace{0.5cm} \text{if } h <0
\;.
\end{equation}
With these definitions, $\mathU(\btheta)$ lives in a space made of $\frac{\Nsites}{2}$
unitary $2\times 2$ matrices. 
This implies 
\[ 
\dim_U = \frac{\Nsites}{2} 4 = 2 \Nsites \;.
\]
To calculate the dimension of gauge degrees of freedom, we need to calculate $\mathGamma$ in this formalism. 
Using the definition of $\mathGamma$ in Eq.~\eqref{eqn:gammadefinition} (i.e. $\mathGamma_{jj'} =  \langle \psi_0| \opbfPhidag{j} \opbfPhi{j'} | \psi_0\rangle$) and the definition of $\opbfPhi{}\!\!$
we find:
\begin{equation}
     \mathGamma = \begin{pmatrix}
\mathbf{D} & 0 \\
0 & \mathbf{D'}
\end{pmatrix},
\end{equation}
with $\mathbf{D}=\operatorname{diag}(0,1,0,1,\dots)$, and
$\mathbf{D'}=\operatorname{diag}(1,0,1,0,\dots)$.
The matrix 
\[ \mathbb{W} =\left( \begin{array}{cc} \mathbf{W} & \mathbf{0} \\
\mathbf{0} & \mathbf{W^*} \end{array}
\right) \;,
\]
must be $2 \times 2$, unitary, block diagonal to preserve the permutational symmetry and must commute with $\mathbb{\Gamma}$ to leave the mean final energy unaltered, see Eqs.~\eqref{eqn:gauge_commut}-\eqref{eqn:unaltereddenergy}. 
If we take a $2\times 2$ unitary matrix:
\[
    \left( \begin{array}{cc} a & b \\
        c & d \end{array} 
\right)
\]
and we impose that it commutes with the blocks
of $\mathGamma$, i.e., with
$\left(\begin{array}{cc} 0 & 0 \\
        0 & 1 \end{array}\right)$
and
$\left(\begin{array}{cc} 1 & 0 \\
        0 & 0 \end{array}\right)$,
%
we find that $c = b = 0$. 
Therefore we can conclude that the $2\times 2$ blocks that compose $\mathbb{W}$ are actually diagonal, and since $\mathbb{W}$ is unitary each element of the diagonal must be a phase. The result is that our gauge freedom is represented by:
\begin{equation} 
\mathbf{W} = 
\operatorname{diag} \left( \nep^{i w_1}, \nep^{i w_2}, \dots, \nep^{i w_{\Nsites}} \right) \;.
\end{equation}
Therefore, $\dim_W = \Nsites$, and  
\begin{equation}
\dim_{U} - \dim_{W} = 2\Nsites - \Nsites = \Nsites \;.
\end{equation} 
As a consequence, the critical number of QAOA layers for the translationally invariant Ising model is predicted to be:
\begin{equation}
\Ptrot^{\mathrm{cr}}_{\Nsites} = \frac{\dim_{U} - \dim_{W}}{2}
= \frac{\Nsites}{2} \;,
\end{equation}
consistently with the results derived in Ref.~\cite{mbeng_quantum_2019}.

\end{document}